\documentclass[a4paper,12pt]{article}

\usepackage{amsmath,amssymb,amscd,mathtools,array,empheq}
\usepackage{lmodern,newtxtext,newtxmath}
\usepackage[a4paper,vmargin=3truecm,hmargin=3truecm]{geometry}
\usepackage[unicode, colorlinks=true, urlcolor=blue, citecolor=blue, linkcolor=blue]{hyperref}
\usepackage{graphicx}
\usepackage{color}
\usepackage[dvipsnames]{xcolor}
\usepackage[english]{babel}
\usepackage{bm}
\usepackage{setspace}
\usepackage{cite}
\usepackage{etoolbox}
\apptocmd{\thebibliography}{\setlength{\itemsep}{0pt}}{}{}
\usepackage[shortlabels]{enumitem}
\numberwithin{equation}{section}

\usepackage{tikz}
\usetikzlibrary{shapes, arrows, arrows.meta, matrix, decorations.markings, calc, babel, quotes, angles, math, patterns}
\makeatletter \g@addto@macro\@floatboxreset\centering \makeatother

\DeclareMathOperator{\Tr}{Tr}
\DeclareMathOperator{\Pf}{Pf}

\DeclareMathOperator{\adj}{adj}

\DeclareMathOperator{\const}{const}
\DeclareMathOperator{\Vect}{Vect}
\DeclareMathOperator{\Diff}{Diff}
\DeclareMathOperator{\Met}{Met}
\DeclareMathOperator{\Geom}{Geom}

\newcommand{\R}{\mathbb{R}}

\newcommand{\Id}{\mathbb{I}}

\newcommand{\vol}{\mathrm{vol}}

\newcommand{\half}{\frac{1}{2}}

\newcommand{\cC}{\mathcal{C}}

\newcommand{\cO}{\mathcal{O}}
\newcommand{\cP}{\mathcal{P}}

\newcommand{\cT}{\mathcal{T}}

\newcommand{\bx}{{\boldsymbol{x}}}
\newcommand{\by}{{\boldsymbol{y}}}

\usepackage{scalerel}
\let\savewidetilde\widetilde
\def\widetilde#1{%
 \ThisStyle{\savewidetilde{\phantom{\SavedStyle#1}}%
  \setbox0=\hbox{$\SavedStyle#1$}\kern-\wd0#1}}

\title{Equations of motion for dynamical systems with angular momentum on Finsler geometries}
 
\author{Lo\"ic Marsot\footnote{marsot3@mail.sysu.edu.cn}, Yuzhang Liu\footnote{Corresponding author: liuyzh58@mail2.sysu.edu.cn},   \\[1.em]
{\normalsize School of Science, Shenzhen campus of Sun Yat-sen University,} \\
{\normalsize No. 66, Gongchang Road, Guangming District, Shenzhen, Guangdong 518107,}\\
{\normalsize P.R. China}}%

\date{{\footnotesize (\today)}}

\begin{document}

\maketitle

\begin{abstract}
This article aims to derive equations of motion for dynamical systems with angular momentum on Finsler geometries. To this end, we apply Souriau's Principle of General Covariance, which is a geometrical framework to derive diffeomorphism invariant equations of motion. The equations we obtain are the generalization of that of Mathisson-Papapetrou-Dixon (MPD) on Finsler geometries, and we give their conserved quantities which turn out to be formally identical to the Riemannian case.

These equations share the same properties as the MPD equations, and we mention different choices possible for supplementary conditions to close this system of equations. These equations have an additional requirement, which is the choice of what defines the tangent direction of the Finsler manifold, since the velocity and momentum are not parallel in general. After choosing the momentum as the Finsler direction and to define the center of mass, we give the complete equations of motion in 3 spatial dimensions, which we find coincide to previously known equations for Finsler spinoptics, and novel equations in 4 dimensions for massive and massless dynamical systems.
\end{abstract}

\section{Introduction}

Finding equations of motion of a dynamical system usually amounts to consider the body as an infinitesimally small test particle of mass $m$ and posit that its trajectory is a geodesic on a suitable manifold. However, in some cases this is only an approximation (albeit a very good one), as some bodies are not infinitesimally small, and in general even elementary particles are usually not only described by their mass, but have some (intrinsic) angular momentum. 

Such dynamical systems with angular momentum, or dipole effects, do not follow geodesics, as the dipole moment may couple to the curvature of spacetime (or to the electromagnetic field if relevant). On (pseudo-)Riemannian geometries, there is a long history of working out equations of motion for such systems. It has culminated in the so-called Mathisson-Papapetrou-Dixon (MPD) equations \cite{Frenkel26,Mathisson37,Papapetrou51,Pirani56,Tulczyjew59,Dixon70},
\begin{subequations}
\label{mpd}
\begin{align}
\frac{\nabla P^\mu}{d\tau} & = - \half {R^\mu}_{\rho\lambda\sigma} S^{\lambda\sigma} \dot{X}^\rho, \label{mpd_pdot}\\
\frac{\nabla S^{\mu\nu}}{d\tau} & = P^\mu \dot{X}^\nu - P^\nu \dot{X}^\mu, \label{mpd_sdot}
\end{align}
\end{subequations}
where $P$ describes the momentum of the system, and $S$ its dipole moment. Note that these equations are not closed, in that an equation for the velocity $\dot{X}$ is missing. Closing this system of equations requires adding supplementary conditions, which amount to choose the center of mass of the system. See \textit{e.g.} \cite{CostaN14} for an extensive discussion on this topic.

These equations are used to compute the trajectory of any kind of spinning test particle in General Relativity, see \textit{e.g.} \cite{PlyatskoSF11,BiniGV17,SkoupyL21,LadinoBLRA23}, gravitational birefringence of light \cite{DuvalS16,DuvalMS18}, and also seem to arise in the scattering of spinning black holes in the post-Minkowskian approximation \cite{Vines18,ChenCHK22}.

These equations are not the only equations predicting a deviation to geodesics for dynamical systems with angular momentum. Taking the example of photons, which are particularly important in Cosmology, as the main object through which observations are carried, and which have spin 1, several independent approaches \cite{GosselinBM06,OanceaJDRPA20} predict the same deviation angle at infinity during gravitational lensing as the MPD equations \cite{DuvalMS18}. In particular, the deviation depends on the helicity (as well as the wavelength) of the photon, leading to \emph{gravitational birefringence}. While the effect is undeniably small, as the birefringence angle of photons passing near the Sun is predicted to be around 8 orders of magnitude lower than the experimental upper bound set in 1974 \cite{Har74},  recent studies have found that more controlled experiments involving light with orbital angular momentum may increase the effect by several orders of magnitude \cite{LianZ24}.

The angular momentum dependency of trajectories is known in other fields as well, for example in optics, it is experimentally known that the polarization of light has an effect on its trajectory, in particular to the so-called Spin Hall Effect of Light \cite{Imbert72,BliokhNKH08,HostenK08}.

\medskip

In parallel, not all dynamical systems are suitably described on (pseudo-)Riemannian manifold, as some fields consider dynamical systems on spaces which are not isotropic. The most well-known example being perhaps optics, where the speed of light depends on its \emph{direction} inside some crystals. There are other examples of fields where the properties of the space depends on the direction one is facing, such as in mechanics \cite{Clayton15}, in mathematical biology and ecology \cite{AntonelliIM93}, and also in cosmology where it has been suggested that dark matter and dark energy could be encoded as such anisotropic degrees of freedom, see \textit{e.g.} \cite{ChangL08} and \cite{HamaHSS21}.

The suitable geometry to describe such dynamical systems is the Finsler geometry. It is a natural extension of Riemannian geometry, where the metric can depend on a direction as well as position. In short, the metric is defined on the tangent bundle $TM$\footnote{Strictly speaking, the metric is better defined on a slightly different space which we will introduce in section \ref{s:finsler}.} instead of the manifold $M$. For the optics example we mentioned earlier, the refractive index in such crystals depends both on the position $\bx$ and on the direction $\by$ the light is going: $n(\bx, \by)$, and so would the metric describing the trajectories in the corresponding manifold by Fermat's principle \cite{Ingarden96,DuvalHH07,Duval08}. 

While the applications of Finsler geometry are diverse, its applications in gravity are especially well developed, from field dynamics considerations \cite{PfeiferW11b}, to finding specific Finsler spacetimes, for example cosmological Finsler spacetimes \cite{HohmannPV20}, or the generalization of a Schwarzschild spacetime \cite{LiC14}, and considering the effect of such Finsler degrees of freedom on the dynamics of photons in these spacetimes \cite{Shen23}.

\medskip

The natural question, and focus of this article, which arises from the previous two points is then: what happens to dynamical systems with dipole moments on Finsler geometries? In other words, can we generalize Finsler geodesics by writing the equivalent of the MPD equations on Finsler geometries?

\medskip

This article is organized as follows. Firstly, in section \ref{s:finsler} we will review the basic elements of Finsler geometry that are used through this article, fixing our notations at the same time. Then, in section \ref{s:pgc} we will quickly review the history of equations of motion and General Covariance in General Relativity, and how Souriau geometrized the situation, and we will show how it can be applied to Finsler geometry to obtain geodesics. Section \ref{s:FMPD} is the main part of the article, in that this is where we apply the PGC to the worldline of dynamical systems with dipole moments to obtain the Finsler-MPD (FMPD) equations, and the conserved quantities associated to these equations. Section \ref{s:supp_cond} is a short section that mentions the role of \emph{supplementary conditions} to close the system of equations. In section \ref{s:spinoptics} we specify our FMPD equations to the case of 3 dimensions using some specific supplementary conditions, in order to recover known equations of Finsler spinoptics. Lastly, in section \ref{s:eom_4d} we specialise our equations to 4 dimensions to obtain novel Finsler equations of motion in spacetime, for both massive and massless systems. 

\section{Some elements of Finsler geometry}
\label{s:finsler}

In this section, which is mostly based on \cite{BaoCS00,ShenS16}, we recall some results of Finsler geometry which we will use through our article. This is also the opportunity to fix our notations, which mostly follow that of the previous references.

A Finsler geometry is defined as a manifold $M$ endowed with a \emph{Finsler structure} $F: TM \to [0, \infty)$, such that,
\begin{enumerate}[i)]
\item $F$ is smooth on $TM_0 = TM \setminus \{0\}$;
\item $F(\bx, \lambda \by) = \lambda F(\bx, \by)$, for $\bx \in M$ and $\by \in T_xM$, and for all $\lambda > 0$;
\item The Hessian 
\begin{equation}
\label{def_finsler_metric}
(g_{ij}) = \half \left(\frac{\partial^2 F^2}{\partial y^i \partial y^j}\right)
\end{equation}    
is positive definite on $TM_0$.
\end{enumerate}

A slightly more general definition, which is relevant in the context of this paper, is to consider a smooth function $L$, homogeneous of degree 2, such that the metric is $(g_{\mu\nu}) = \half \left(\frac{\partial^2 L}{\partial y^\mu \partial y^\nu}\right)$ \cite{PfeiferW11b,ShenS16}. If the metric is positive definite, then the two definitions are equivalent with $L = F^2$. However the latter definition allows for a more general $g$ with the possibility of negative eigenvalues, which is relevant for space-time considerations where Lorentzian signature is required.

A seemingly natural basis for tensors on $TM_0$ is $\left\lbrace\frac{\partial}{\partial x^\mu}, \frac{\partial}{\partial y^\mu}, dx^\mu, dy^\mu\right\rbrace$. However, it turns out that $\frac{\partial}{\partial x^\mu}$ and $dy^\mu$ do not transform linearly under a coordinate change on $TM$ induced by $M$, since for a transformation $x^\mu = x^\mu(\tilde{x}^1, \ldots, \tilde{x}^n)$ one has \cite{BaoCS00} $\dfrac{\partial}{\partial \tilde{x}^\mu} = \dfrac{\partial x^\lambda}{\partial \tilde{x}^\mu} \dfrac{\partial}{\partial x^\lambda} + \dfrac{\partial^2 x^\lambda}{\partial \tilde{x}^\mu \partial \tilde{x}^\sigma} \tilde{y}^\sigma \dfrac{\partial}{\partial y^\lambda}$ where $y^\mu = \dfrac{\partial x^\mu}{\partial \tilde{x}^\lambda} \tilde{y}^\lambda$, and similarly for $dy^\mu$.

To remedy this, note that a choice of $F$ induces a splitting of $T(TM_0)$, where the horizontal part is spanned by 
\begin{equation}
\frac{\delta}{\delta x^\mu} = \frac{\partial}{\partial x^\mu} - N^\nu{}_\mu \frac{\partial}{\partial y^\nu}
\end{equation}
where $N^\nu{}_\mu$ are the coefficients of the \emph{non-linear connection}, 
\begin{equation}
N^\nu{}_\mu = \half \frac{\partial G^\nu}{\partial y^\mu},
\end{equation}
with $G^\nu$ the \emph{spray coefficients},
\begin{align}
\label{spray_coef}
G^\nu & = \half g^{\nu\lambda} \left(\frac{\partial^2 F^2}{\partial y^\lambda \partial x^\sigma} y^\sigma - \frac{\partial F^2}{\partial x^\lambda}\right)
\end{align}
while the vertical part is spanned by the $\partial/\partial y^\mu$. The same is true for $T^*(TM_0)$, where the horizontal part is spanned by $dx^\mu$ and the vertical part by 
\begin{equation}
\delta y^\mu = dy^\mu + N^\mu{}_\nu dx^\nu
\end{equation}
Hence, $\frac{\delta}{\delta x^\mu}$ and $\delta y^\mu$ transform linearly under coordinate change. Since they are dual to, respectively $dx^\mu$ and $\frac{\partial}{\partial y^\mu}$, in the sense that $dx^\mu(\delta/\delta x^\nu) = \delta^\mu_\nu$ and $\delta y^\mu(\partial/\partial y^\nu) = \delta^\mu_\nu$, the set $\left(\dfrac{\delta}{\delta x^\mu}, \dfrac{\partial}{\partial y^\mu}, dx^\mu, \delta^\mu \right)$ forms a suitably covariant basis for tensors on $TM_0$.

\medskip

The Finsler structure $F$ being positively homogeneous of degree 1 has some rather important consequences for us. The main effect is that one can work on a space where all objects of interest are invariant under rescaling of $y$. For example, the metric \eqref{def_finsler_metric} is invariant under such rescalings, as well as the Cartan tensor $A_{\mu\nu\lambda} = \frac{F}{2} \frac{\partial g_{\mu\nu}}{\partial y^\lambda}$. The tangent of the tangent bundle is too big of a space to accommodate these invariant objects, and it makes more sense to work with vectors on $\pi^* TM_0$, where $\pi: TM_0 \rightarrow M$ is the natural projection, \textit{i.e.} $\pi(x, y) = x$, rather than on $T(TM_0)$. The most notable example being the distinguished section $l = l^\mu \partial_{x^\mu}$, where $l^\mu = y^\mu/F$. In short, $\pi^* TM_0$ is made of a copy of $T_xM$ on top of each point in $TM_0$. See \textit{e.g.} \cite[Chap. 2]{BaoCS00} or \cite{ShenS16} for more detailed explanations.

Then, for example, the metric is the bilinear symmetric form $g: \pi^*TM \times \pi^*TM \rightarrow \R$, such that
\begin{equation}
\label{metric_pitm}
g = g_{\mu\nu}(x, y) dx^\mu \otimes dx^\nu = \half \frac{\partial^2 F^2}{\partial y^\mu \partial y^\nu} dx^\mu \otimes dx^\nu
\end{equation}

Moreover, since the metric $g$ is homogeneous of degree 0, Euler theorem immediately implies that the Cartan tensor, which is symmetric in all its indices, contracted with the vector field $l$ vanishes,
\begin{equation}
A_{\mu\nu\lambda} l^\mu = 0
\end{equation}

Unlike in Riemannian geometry, there is no canonical affine connection on $\pi^* TM$, since it is impossible to have an affine connection that is both compatible with the metric $g$, and torsionless. However several notable examples exist, in particular for us the Chern connection, which is torsionless and \emph{almost} compatible with the metric; or the Cartan connection which is compatible however has torsion. We will differentiate between objects derived from these two connections by having a hat on all objects derived from the Cartan connection $\widehat{\omega}$. Other choices of connection are also possible, depending on the use case.

The Chern connection $\omega_\nu{}^\mu = \Gamma^\mu_{\nu\lambda} dx^\lambda$,
\begin{equation}
\Gamma^\mu_{\nu\lambda} = \half g^{\mu\sigma} \left(\frac{\delta g_{\sigma\lambda}}{\delta x^\nu} + \frac{\delta g_{\nu\sigma}}{\delta x^\lambda} - \frac{\delta g_{\nu\lambda}}{\delta x^\sigma}\right)
\end{equation}
is torsionless, and almost $g$-compatible, in the sense that the covariant derivative $\nabla: \Gamma(\pi^* TM_0) \rightarrow \Gamma(T(TM_0)) \otimes \Gamma(\pi^*TM)$ is such that
\begin{equation}
\label{nabla g chern}
(\nabla g)_{\mu\nu} = \left(dg_{\mu\nu} - \omega_\mu{}^\lambda g_{\lambda\nu} - \omega_\nu{}^\lambda g_{\mu\lambda}\right) = 2 A_{\mu\nu\rho} \frac{\delta y^\rho}{F}
\end{equation}
We can decompose the covariant derivative in terms of an horizontal and a vertical derivative. If $T$ is a section of $\pi^*TM \otimes \pi^*T^*M$, 
\begin{equation}
\label{cov_der}
\nabla T^\mu{}_\nu = T^\mu{}_\nu{}_{|\lambda} dx^\lambda + T^\mu{}_\nu{}_{;\lambda} \frac{\delta y^\lambda}{F}
\end{equation}
where
\begin{align}
T^\mu{}_\nu{}_{|\lambda} & = \frac{\delta T^\mu{}_\nu}{\delta x^\lambda} + T^k{}_\nu \Gamma^\mu_{\sigma\lambda} - T^j{}_k \Gamma^\sigma_{\nu\lambda} \\
T^\mu{}_\nu{}_{;\lambda} & = F \frac{\partial T^\mu{}_\nu}{\partial y^\lambda}
\end{align}

From \eqref{nabla g chern} we see that the covariant derivative of the metric, using the Chern connection, is purely vertical, in the sense that $g_{\mu\nu|\lambda} = 0$ and $g_{\mu\nu;\lambda} = 2 A_{\mu\nu\lambda}$. The covariant derivative of the distinguished section $l$ is noteworthy as well, as we will make use of $l^\mu{}_{|\lambda} = 0$ and $l^\mu{}_{;\lambda} = \delta^\mu_\lambda - l^\mu l_\lambda$.

The curvature tensor of the Chern connection is then $\Omega_\nu{}^\mu = d\omega_\nu{}^\mu - \omega_\nu{}^\lambda \wedge \omega_\lambda{}^\mu = \half R_\nu{}^\mu{}_{\lambda\sigma} dx^\lambda \wedge dx^\sigma + \cP_\nu{}^\mu{}_{\lambda\sigma} dx^\lambda \wedge \frac{\delta y^\sigma}{F}$, where
\begin{align}
R_\nu{}^\mu{}_{\lambda\sigma} & = \frac{\delta \Gamma^\mu_{\nu\sigma}}{\delta x^\lambda}  - \frac{\delta \Gamma^\mu_{\nu\lambda}}{\delta x^\sigma}  + \Gamma^\mu_{\kappa\lambda} \Gamma^\kappa_{\sigma\nu} - \Gamma^\mu_{\kappa\sigma} \Gamma^\kappa_{\lambda\nu} \\
\cP_\nu{}^\mu{}_{\lambda\sigma} & = - F \frac{\partial \Gamma^\mu_{\nu\lambda}}{\partial y^\sigma}
\end{align}
Note that the curvature could in principle contain a term of the form $\half Q_\nu{}^\mu{}_{\lambda\sigma} \frac{\delta y^\lambda}{F} \wedge \frac{\delta y^\sigma}{F}$, however this term turns out to vanish with the Chern connection.

Later on, we will make use of the following formulas for the commutation of derivatives,
\begin{align}
\label{interchangehh}
T^\mu{}_{\nu|\sigma|\lambda} - T^\mu{}_{\nu|\lambda|\sigma} & = T^\kappa{}_\nu R_\kappa{}^\mu{}_{\lambda\sigma} - T^\mu{}_\kappa R_\nu{}^\kappa{}_{\lambda\sigma} - T^\mu{}_{\nu;\kappa} R_\gamma{}^\kappa{}_{\lambda\sigma} l^\gamma \\
\label{interchangehv}
T^\mu{}_{\nu;\sigma|\lambda} - T^\mu{}_{\nu|\lambda;\sigma} & = T^\kappa{}_\nu \cP_{\kappa}{}^\mu{}_{\lambda\sigma} - T^\mu{}_\kappa \cP_\nu{}^\kappa{}_{\lambda\sigma} + T^\mu{}_{\nu;\kappa} A^\kappa{}_{\lambda\sigma|\gamma} l^\gamma
\end{align}

The Cartan connection $\widehat{\omega}_\nu{}^\mu$ is related to the Chern connection by
\begin{equation}
\label{link_cartan_chern}
\widehat{\omega}_\nu{}^\mu = \omega_\nu{}^\mu + A^\mu{}_{\nu\lambda} \frac{\delta y^\lambda}{F}
\end{equation}
We immediately see from \eqref{nabla g chern} and \eqref{link_cartan_chern} that the Cartan covariant derivative $\widehat{\nabla}$ is compatible with the metric, $\widehat{\nabla} g = 0$. The drawback of this connection is that its torsion does not vanish, $\widehat{\Omega}^\mu = - A^\mu{}_{\nu\lambda} dx^\nu \wedge \frac{\delta y^\lambda}{F}$, although we will not make use of this fact.

The Cartan curvature is also decomposed as $\widehat{\Omega}_\nu{}^\mu = \half \widehat{R}_\nu{}^\mu{}_{\lambda\sigma} dx^\lambda \wedge dx^\sigma + \widehat{\cP}_\nu{}^\mu{}_{\lambda\sigma} dx^\lambda \wedge \frac{\delta y^\sigma}{F} + \half \widehat{Q}_\nu{}^\mu{}_{\lambda\sigma} \frac{\delta y^\lambda}{F} \wedge \frac{\delta y^\sigma}{F}$, where each tensors is related to their Chern counterpart by \cite{BaoCS96},
\begin{align}
\widehat{R}_\nu{}^\mu{}_{\lambda\sigma} & = R_\nu{}^\mu{}_{\lambda\sigma} + A^\mu{}_{\nu\rho} R_\kappa{}^\rho{}_{\lambda\sigma} l^\kappa \\
\widehat{\cP}_\nu{}^\mu{}_{\lambda\sigma} & = \cP_\nu{}^\mu{}_{\lambda\sigma} + A^\mu{}_{\nu\sigma|\lambda} - A^\mu{}_{\nu\rho} A^\rho{}_{\lambda\sigma|\kappa} l^\kappa \\
\widehat{Q}_\nu{}^\mu{}_{\lambda\sigma} & = - 2 A^\kappa{}_{\nu[\lambda} A^\mu{}_{\sigma]\kappa}
\end{align}
Note that the difference $\widehat{R}_{\nu\mu\lambda\sigma} - R_{\nu\mu\lambda\sigma}$ is symmetric in $\nu$ and $\mu$, and likewise for $\cP$. In this article, these objects will most often appear contracted with a skewsymmetric tensor $S^{\nu\mu}$, and so we will have that $\widehat{R}(S)_{\lambda\sigma} = \widehat{R}_{\nu\mu\lambda\sigma} S^{\nu\mu} = R(S)_{\lambda\sigma}$ \cite{Duval08}, and likewise for $\cP(S)$, whose expression is,
\begin{equation}
\label{ps}
\cP(S)_{\mu\nu} = \cP_{\lambda\sigma\mu\nu} S^{\lambda\sigma} = 2 \left(A_{\mu\nu\lambda|\sigma} - A_{\lambda\mu\kappa} A^\kappa{}_{\sigma\nu|\rho} l^\rho\right)S^{\lambda\sigma}
\end{equation}
The definition of $\widehat{Q}(S)$ is similar, though it is obviously not equal to $Q(S)$,
\begin{equation}
\label{qs}
\widehat{Q}(S)_{\mu\nu} = \widehat{Q}_{\lambda\sigma\mu\nu} S^{\lambda\sigma} = - 2 A_{\lambda\kappa\mu} A^\kappa{}_{\sigma\nu} S^{\lambda \sigma}
\end{equation}

\medskip

Since we aim to study trajectories of dynamical systems, we will deal with curves on $M$. In Finsler geometry, the tangent component $y$ is usually not independent of the behavior of the trajectory on $M$. Most commonly $y^\mu = \frac{dx^\mu}{d\tau}$ is set. Because of this, a transformation on $x$ also affects a transformation on $y$. Hence, if we have a vector field $\xi = \xi^\mu(x) \frac{\partial}{\partial x^\mu} \in \Vect(M)$, to consider its action on an object in $TM_0$, we will need to consider the complete (or natural) lift \cite{YanoK66} of this vector field to $\bar{\xi} \in \Vect(TM_0)$,
\begin{equation}
\label{natural_lift_xi}
\bar{\xi} = \xi^\mu \frac{\partial}{\partial x^\mu} + \frac{\partial \xi^\mu}{\partial x^\nu} y^\nu \frac{\partial}{\partial y^\mu} 
\end{equation}
such that $\bar{\xi}$ is the generator of the tangent map $T\Phi_t$ to the flow $\Phi_t$ induced by $\xi$ on $M$, with $T\Phi_t(x, y) = (\Phi_t(x), (d\Phi_t)_x(y))$.

The directional derivative along a curve $C : \tau \mapsto (X^\mu(\tau), y^\mu(\tau))$ on $TM_0$ is
\begin{align}
\frac{d}{d\tau} & = \frac{dx^\mu}{d\tau} \frac{\partial}{\partial x^\mu} + \frac{dy^\mu}{d\tau} \frac{\partial}{\partial y^\mu} \\
& = \dot{X}^\mu \frac{\delta}{\delta x^\mu} + \dot{Y}^\mu \frac{\partial}{\partial y^\mu}
\end{align}
where we set $\dot{Y}^\mu = \frac{dy^\mu}{d\tau} - N^\mu{}_\nu \dot{X}^\nu$

Tensors describing the state of the dynamical system, \textit{e.g.} the momentum tensor $P$, belong to $\pi^*TM_0$. The covariant derivative of such objects along a curve parametrized by $\tau$ is then
\begin{equation}
\label{def_cov_der_curve}
\frac{\nabla}{d\tau} P^\mu = \nabla_{(\dot{X},\dot{Y})} P^\mu =  \dot{X}^\rho P^\mu{}_{|\rho} + \dot{Y}^\rho P^\mu{}_{;\rho}
\end{equation}

Lastly, we will be interested in the variations of metrics along some direction generated by $\xi$ on $M$, hence we need the expression of the Lie derivative. However since metrics are objects on $\pi^*TM$, we actually need to consider the variations along the natural lift $\bar{\xi}$ \eqref{natural_lift_xi} of $\xi$, \cite{Davies39}, 
\begin{equation}
\label{lie_der}
L_{\bar{\xi}} g_{\mu\nu} = \xi_{\mu|\nu} + \xi_{\nu|\mu} + 2 A^\lambda{}_{\mu\nu} l^\sigma \xi_{\lambda|\sigma}
\end{equation}

\section{Geodesics from diffeomorphism invariance}
\label{s:pgc}

\subsection{The Principle of General Covariance}

The question of where do the equations of motion come from in General Relativity (GR) is a long-standing one, already from Einstein. There were some attempts to derive equations of motion for elementary particles from the fields equations, see \textit{e.g.} the famous Einstein-Infeld-Hoffman approach \cite{EinsteinIH38}. Another approach, initially split in two seemingly different approaches, is based on the principle of \emph{General Covariance} (GC), which is at the heart of GR, which states that the physics should be invariant under change of coordinates. A more ``active'' statement would be that objects in GR should be invariant under diffeomorphisms. This latter approach has turned out very successful, and it is the one we will use to develop the equations of motion for Finsler geometry. Let us present it in more details in this section.

While in General Relativity the object of choice to work with is the metric $g$, which is an object belonging to the space of all metrics $\Met(M)$ (of a given signature), the principle of General Covariance deals with the redundancy of the metric: any metric $g'$ obtained from the first one by a diffeomorphism produces an equivalent physical description. In a sense this makes metrics awkward to use, and one would rather want to work with objects invariants under diffeomorphisms, which map one-to-one with the physical information of the Universe, with the hope that such invariant objects lead to natural equations or statements. These invariant objects belong to what we will call the space of Geometries, in line with \cite{Souriau74}.

The mathematical structure of this space of Geometries, which is essentially identified with the quotient $\Met(M)/\Diff(M)$ (up to some technical considerations), has been studied in \cite{Ebin70}. This kind of bundle geometry is not a lone example, for example a few years later, Singer studied in \cite{Singer78} the geometry of the set of all connections seen as the bundle with structure group the gauge group of these connections to study the Gribov ambiguity. In general, studying such invariant objects is insightful, see \textit{e.g.} \cite{Francois24,FrancoisR25} for more details and a method to built such objects in general.

\medskip

In parallel to these geometric considerations, in 1937 Mathisson developed a much more practical framework \cite{Mathisson37} (\cite{Mathisson10} for an english translation) to study the equations of motion for extended bodies, \textit{i.e.} objects with an internal structure. See \cite{Dixon15} for an excellent review of Mathisson's and subsequent work. In short, this approach works by considering the extended body as a point particle, at a position to be chosen (not uniquely), and to describe its internal structure by writing its energy-momentum tensor on its worldline $\cC$. The energy-momentum tensor can then be expanded as the sum of its multipoles, the monopole describing essentially the linear momentum, or mass, of the body, the dipole its angular momentum, and so on. Then, the conservation of the energy-momentum tensor should lead to equations of motion.

More explicitly, he wrote, for $T$ the energy momentum tensor and $\varphi$ a test 2-tensor with compact support,
\begin{equation}
\label{mathisson_functionals}
\int T^{\mu\nu} \varphi_{\mu\nu} \sqrt{-g} d^4 x = \int_\cC \left(m^{\mu\nu} \varphi_{\mu\nu} + m^{\lambda\mu\nu} \nabla_\lambda \varphi_{\mu\nu} + \ldots \right) ds
\end{equation}
where instead of evaluating the left-hand side on the whole spacetime, the multipoles $(m^{\mu\nu}, m^{\lambda\mu\nu}, \ldots)$ are evaluated on the expansion of $\varphi$ on the worldline of the particle. In a more modern language (that did not exist at his time), we say this is a distribution $\cT(\varphi)$.

The conservation of the energy momentum tensor $\nabla_\mu T^{\mu\nu} = 0$ is then equivalent to the vanishing of the left-hand side of the above expression with $\varphi_{\mu\nu} = \nabla_{(\mu} \xi_{\nu)}$ for some vector $\xi$. The equations of motion are then given by the right-hand side vanishing for that expression of $\varphi$,
\begin{equation}
\int_\cC \left(m^{\mu\nu} \nabla_{(\mu} \xi_{\nu)} + m^{\lambda\mu\nu} \nabla_\lambda \nabla_{(\mu} \xi_{\nu)} + \ldots \right) ds = 0
\end{equation}
Usually, one considers small enough bodies that the higher multipoles can be neglected, so that we can only consider the monopole, or monopole and dipole moments. If one considers only the monopole moment, Mathisson showed that $\int_\cC m^{\mu\nu} \nabla_{(\mu} \xi_{\nu)} ds = 0 \, \forall \xi$ leads to $m^{\mu\nu} = M v^\mu v^\nu$ where $v$ is the unit tangent vector to the worldline, and that $\cC$ must be a geodesic. Then, if one keeps both monopole and dipole, the above equation leads to the now-called MPD equations \eqref{mpd}, although this explicit form was shown in later work. There is much more to his work, however in this present paper we are concerned about the MPD equations and the energy-momentum tensor distribution.

\medskip

Then, in 1974\footnote{Souriau already published a note about this in 1970 \cite{Souriau70b}, but it is very short, with few details, and no citations except his own book.} Souriau recast Mathisson's idea in the context of diffeomorphism invariance. The key idea is that the conservation of the energy-momentum tensor follows from diffeomorphism invariance, since the special form of the test tensor $\varphi_{\mu\nu} = \nabla_{(\mu} \xi_{\nu)}$ in Mathisson's work is none other than the Lie derivative of the metric by $\xi$, \textit{i.e.} an infinitesimal diffeomorphism.

While this seems like minor progress, Souriau gave a complete geometric picture (following the $\Met(M)/\Diff(M)$ idea of Ebin, although it is unclear if Souriau knew about it since \cite{Ebin70} is not cited.) of the derivation of these MPD equations, together with an elegant way to obtain the conserved quantities. His description being based on purely geometric terms also allowed him to include electromagnetism to the MPD equations, and will allow us in the present article to generalize the MPD equations to Finsler geometry. Hence, in the rest of this subsection, we review his geometric description.

First, note that this derivation of equations of motion starts from finding an energy momentum tensor satisfying diffeomorphism invariance, rather than the more usual search of a Lagrangian. These methods are not equivalent in general, and it seems that starting from an energy momentum tensor is slightly more general, because in the case of the MPD equations it is impossible to find a Lagrangian defined on the same configuration space as the energy momentum tensor due to topological reasons\footnote{Because the symplectic 2-form of $S^2$, to describe the spin, admits no global potential.} \cite{Souriau74}. Although it would be possible if one enlarges the configuration space \cite{DamourI24}. Nevertheless in some cases the two methods are equivalent \cite{Iglesias19}.

\medskip

We now review the geometric framework as described in \cite{Souriau74}, which starts by the consideration of $\Met(M)/\Diff_c(M)$. Recall that from \cite{Ebin70}, the physical information is encoded not in the space of metrics but on its quotient by diffeomorphisms, the space of geometries, which we denote by $\Geom(M)$. Now, the quotient is not taken by all diffeomorphisms, as if one keeps isometries, the quotient would fail to be a manifold in general. Instead, one considers the diffeomorphisms of $M$ with compact support $\Diff_c(M)$, which isometries are not a part of (if $M$ is well-behaved, for example if two points determine at most one geodesic \cite{Souriau74}). As Souriau showed, the (non-compact) isometries will nevertheless play a role later to obtain conserved quantities, but not on the quotient.  Hence the definition,
\begin{equation}
\label{def_geom}
\Geom(M) = \Met(M) / \Diff_c(M).
\end{equation}

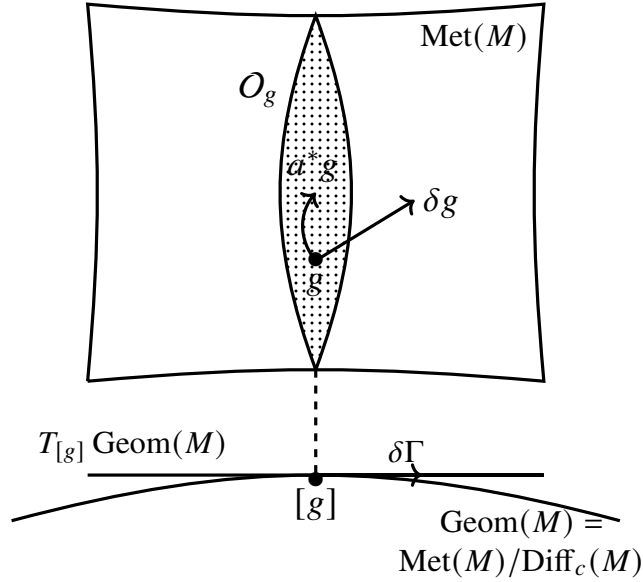
\begin{figure}[ht]
\centering
\begin{tikzpicture}[line width=1.2pt,scale=1, every node/.style={transform shape}]
  \draw (0,0) to[out=5,in=175] coordinate[pos=0.5](midmid) (6,0) to[out=95,in=-95] node[pos=0.92,left,scale=1]{$\Met(M)$} (6,5) to[out=-175,in=-5] coordinate[pos=0.5](topmid) (0,5) to[out=-85,in=85] (0,0);
  \draw (-1,-1.85) to[out=15,in=165] node[pos=0.5,below,scale=1.1]{$[g]$} node[pos=0.85,below,align=center]{$\Geom(M) =$ \\ $\Met(M)/\Diff_c(M)$} coordinate[pos=0.5](botmid) (7,-1.85);
	  \draw [pattern=dots] (midmid) to[out=110,in=-110] node[pos=0.8,left,scale=1.1]{$\, \cO_g$} (topmid) to [out=-70,in=70] (midmid);
  \draw [->] (3,1.6) -- node[pos=0,scale=1.2] {$\bullet$} node[pos=0,scale=1.1,below] {$g$} node [pos=0.95, right,scale=1.1] {$\delta g$} (4.3,2.4);
  \draw [->] (3,1.6) to[out=130,in=-130] node [pos=0.95, above,scale=1.1] {$a^* g$} (3,2.5);
  \draw [dashed] (midmid) -- node [pos=1.05,scale=1.2] {$\bullet$} (botmid);
  \draw (botmid) -- +(3,0) -- node[pos=0.90,above] {$T_{[g]}\Geom(M)$} +(-3,0);
  \draw [->] (botmid) -- node[pos=0.85,above] {$\delta \Gamma$} +(1.4,0);
\end{tikzpicture}
\caption{The geometry of the Principle of General Covariance. The dotted area represents the orbit $\cO_g$ of all metrics connected to the metric $g$ by a diffeomorphism. This orbit is projected as a point $[g]$ on the quotient space $\Geom(M)$.}
\label{f:pgc}
\end{figure}

Now, $\Geom(M)$ is a complicated space, but we do not need to characterize much of its geometry. Consider a point $g \in \Met(M)$, and $\cO_g$ the orbit of $g$ by $\Diff_c(M)$, \textit{i.e.} the set of all metrics one can obtain as $a^* g$ with $a \in \Diff_c(M)$, see figure \ref{f:pgc} for a picture of the situation. We write an infinitesimal variation of the metric $g$ as $\delta g$. If this variation is an infinitesimal diffeomorphism, then it is given by the Lie derivative of the metric by the vector field $\xi \in \Vect_c(M)$ which generates the infinitesimal diffeomorphism, \textit{i.e.} $\delta g = L_\xi g \in T_g\cO_g \subset T_g \Met(M)$.

Now, look at the projection of this variation on $\Geom(M)$, $\delta g \mapsto \delta \Gamma$. By construction, each orbit $\cO_g$ on $\Met(M)$ maps to a point $[g] \in \Geom(M)$. While a general variation $\delta g \in T_g \Met(M)$ maps to some $\delta \Gamma \in T_{[g]} \Geom(M)$, the case of interest is that of a variation generated by an infinitesimal diffeomorphism, \textit{i.e.} $\delta g \in T_g \cO_g$. In that case, it maps (by construction) to a vanishing variation $\delta \Gamma = 0$ on $T_{[g]} \Geom(M)$. In other words, the effects of a diffeomorphism are unobservable: the physical content represented on the geometry is the same, so the point on $\Geom(M)$ does not change. One can also say that $T_g \cO_g$ is the vertical tangent space of $\Met(M)$ at $g$.

Consider next the cotangent space $T^*_g \Met(M)$. It can be seen as the space of linear functionals of $\delta g$. It is on this space that the functionals $\cT$ introduced by Mathisson \eqref{mathisson_functionals} live, before using the conservation of the energy-momentum tensor, for general test tensors $\varphi$. Ultimately, the Principle of General Covariance states that one should not consider energy-momentum functionals on $T^*_g \Met(M)$, but rather on the $\Diff(M)$-invariant $T^*_{[g]} \Geom(M)$. The projection $\cT \in T^*_g \Met(M) \mapsto \widetilde{\cT} \in T^*_{[g]} \Geom(M)$ is such that $\widetilde{\cT}(\delta \Gamma) = \cT(\delta g)$. Given the considerations of the previous paragraph, the left-hand side vanishes when $\delta g$ is an infinitesimal diffeomorphism. Hence, a necessary condition is that $\cT(\delta g) = 0$ for such variations.

In fact, this is all we need to know to state Souriau's version of the Principle of General Covariance: the geometric information of the Universe, as encoded in the energy-momentum tensor, for example that of the worldline of a dynamical system, is represented by a distribution tensor $\cT \in T^*_g \Met(M)$ such that
\begin{equation}
\label{PGC}
\boxed{\cT(L_{\xi} g) = 0, \quad \forall \xi \in \Vect_c(M)}
\end{equation}

This is readily equivalent to the construction of Mathisson: if one writes $\cT(\delta g) = \int_M T^{\mu\nu} \delta g_{\mu\nu} \vol$, an integral by part shows that $\cT(L_\xi g) = 0, \, \forall \xi$ is equivalent ot the conservation of the energy-momentum tensor, $\nabla_\mu T^{\mu\nu} = 0$. However, as stated before, this geometrization allows for a much easier time when generalizing the equations of motion to different spaces. For example, Souriau applies it not only to the space of all metrics, but also to the space of all gravitational and electromagnetic potentials, i.e. the set of $(g, A)$, such that the structure group is the semi direct product of diffeomorphisms and the gauge group, and which leads to the MPD equations in both a gravitational and electromagnetic background.
 
\medskip

In the following, we will consider the generalization of this construction to that of Finsler geometry. The first thing to note is that the above construction is for Riemannian geometry. The case of Finsler geometry is slightly different in that mathematical objects live on $TM_0$ instead of $M$. In particular, integrals should be taken along a curve on $TM_0$. There is a subtle point here: we will \emph{not} consider the integration curve on $TM_0$ to be the natural lift of the worldline on $M$, which would identify $\dot{X} = y$. Instead, we will consider it to be a general curve on $TM_0$. Justification for this will be made in section \ref{s:supp_cond}, but in short, for trajectories with dipole moments, in general $\dot{X} \neq y$. However, the idea behind the principle does not change, and \eqref{PGC} stays formally the same, other than $g$ being defined on $TM_0$ and $\xi$ being replaced by its natural lift $\bar{\xi}$ on $TM_0$.

For people working in Finsler geometry, it might seem more natural to apply Souriau's Principle of General Covariance using the Finsler structure $F$ rather than the metric $g$, since the latter contains some redundancy. However, we choose to use the Principle expressed in terms of $g$ instead of $F$ as we find it more straightforward, in particular for the dipole case in section \ref{s:FMPD}. It is also then easier to compare the obtained Finsler moments with the existing (Riemannian) literature.

Note also that using general covariance in the context of Finsler geometries has already been considered, although for field theories, see \cite{HohmannPV22}.

\subsection{Finsler geodesics}

The distribution tensor $\cT \in T^*_M \Met(M)$ representing the energy-momentum distribution of a particle travels on a worldline $\cC$ on $TM_0$, evaluated on an arbitrary variation $\delta g$ may be written as
\begin{equation}
\label{general_distrib}
\cT(\delta g) = \half \int_\cC f(\delta g, \nabla \delta g, \ldots) d\tau
\end{equation}
for some linear function $f$ of $\delta g$ and its derivatives, $\tau$ a parameter along the worldline, and the $1/2$ factor being here for later convenience. The function $f$ can be written as $f(\delta g, \nabla \delta g, \ldots) = \theta(\delta g) + \Phi(\nabla \delta g) + \ldots$ for some coefficients $\theta, \Phi, \ldots$. 

If one neglects the effect of dipole and higher moments, only the first term of $f$ is kept, and the distribution $\cT$ takes the form,
\begin{equation}
\label{distribution_particle}
\cT(\delta g) = \half \int_\cC \theta^{\mu\nu} \delta g_{\mu\nu} d\tau.
\end{equation}
for a symmetric 2-tensor $\theta$, yet to be determined.

\medskip

We are now ready to apply the Principle of General Covariance \eqref{PGC}. This happens in two steps. First, it is used with some particular vector fields to obtain the expression of $\theta$, and once that is done, the PGC for a general vector field will yield the equations of motion.

First, consider the variation $\delta g = L_{\alpha \xi} g$, where $\alpha \in C^\infty(M)$ is a function that vanishes on the worldline of the particle, $\alpha|_\cC = 0$, and $\xi \in \Vect_C(M)$ any vector field. Note that neither $\alpha$ nor $\xi^\mu$ depend on $y$. We calculate,
\begin{equation}
\half (L_{\alpha \xi} g)_{\mu\nu} = (\partial_{(\mu} \alpha) \xi_{\nu)} + \alpha \xi_{(\mu|\nu)} + A_{\mu\nu\lambda} l^\sigma (\partial_\sigma \alpha) \xi^\lambda + \alpha A_{\mu\nu\lambda} l^\sigma \xi^\lambda{}_{|\sigma}
\end{equation}

Given that the integral \eqref{distribution_particle} is over $\cC$, since $\alpha$ vanishes on $\cC$, and taking into account the symmetry $\theta^{\mu\nu} = \theta^{(\mu\nu)}$, only terms containing derivatives of $\alpha$ do not vanish,
\begin{align}
\cT(L_{\alpha \xi} g) & = \int_\cC \theta^{\mu\nu} \xi_\mu (\partial_\nu \alpha) d\tau + \int_\cC \theta^{\mu\nu} A_{\mu\nu\lambda} l^\sigma (\partial_\sigma \alpha) \xi^\lambda d\tau \\
& = \int_\cC \xi_\mu \left(\theta^{\mu\nu}  + \theta^{\lambda\sigma} A_{\lambda\sigma}{}^\mu l^\nu\right) \partial_\nu \alpha \, d\tau 
\end{align}
Since, by account of the PGC \eqref{PGC}, the above integral vanishes for all $\xi$, $\theta$ must be a solution of,
\begin{equation}
\label{geod_emt_gradalpha}
\left(\theta^{\mu\nu}  + \theta^{\lambda\sigma} A_{\lambda\sigma}{}^\mu l^\nu\right) \partial_\nu \alpha = 0
\end{equation}
By construction, $\alpha$ vanishes on the worldline, which means that its gradient $\partial_\nu \alpha$ is orthogonal to the worldline. In other words, we have $\dot{X}^\nu \partial_\nu \alpha = 0$, where $\dot{X}^\nu = dx^\nu/d\tau$ is the vector field tangent to the worldline.

The equation \eqref{geod_emt_gradalpha} is of the form $\vartheta^{\mu\nu} \partial_\nu \alpha = 0$ when evaluated on the worldline. Since this has to hold true for all choices of functions $\alpha$ that vanish on a given worldline, the only solution is for $\vartheta$ to be proportional to $\dot{X}$ on its second index, \textit{i.e.} it must be of the form $\vartheta^{\mu\nu} = P^\mu \dot{X}^\nu$ for some unknown vector $P$. Hence, this means that \eqref{geod_emt_gradalpha} is equivalent to 
\begin{equation}
\label{particle_pgc_condition}
\theta^{\mu\nu}  + \theta^{\lambda\sigma} A_{\lambda\sigma}{}^\mu l^\nu = P^\mu \dot{X}^\nu
\end{equation}
for some vector $P$ yet to be determined. Note that this equality is only valid on the worldline of the particle.

This equation can be solved for $\theta$ in an iterative way, by replacing $\theta$ by $\theta^{\mu\nu} = P^\mu \dot{X}^\nu - \theta^{\lambda\sigma} A_{\lambda\sigma}{}^\mu l^\nu$ in the second term, and remembering that $A$ contracted with $l$ vanishes,
\begin{equation}
\theta^{\mu\nu} = P^\mu \dot{X}^\nu - P^\lambda \dot{X}^\sigma A_{\lambda\sigma}{}^\mu l^\nu
\end{equation}
However, the above solution for $\theta$ has to be symmetric in $\mu$ and $\nu$ since $\theta$ is. Since $A_{\cdot \cdot}{}^\mu$ is orthogonal to $l^\mu$, the only non trivial possibility is that $P \propto l$, which means that the second term vanishes, which in turns implies $P \propto \dot{X}$. Hence, the solution for $\theta$ is,
\begin{equation}
\label{solution_theta_geodesic}
\theta^{\mu\nu} = P^\mu \dot{X}^\nu, \qquad P \propto \dot{X} \propto l
\end{equation}

When replacing $\theta$ by its solution above and when evaluated on $\delta g = L_{\bar{\xi}} g$ (recall that $\bar{\xi}$ is the natural lift of $\xi$, cf. \eqref{natural_lift_xi}), the distribution \eqref{distribution_particle} for our particle becomes particularly simple,
\begin{equation}
0 = \cT(L_{\bar{\xi}} g) = \int_\cC P^\mu \dot{X}^\nu \xi_{\mu|\nu} d\tau
\end{equation}

We would like to transfer the directional derivative onto $P$ rather than $\xi$. However remember that the covariant derivative along a curve in Finsler geometry contains a term involving $\dot{Y}$, see \eqref{def_cov_der_curve}. We thus have,
\begin{align}
0 = \cT(L_{\bar{\xi}} g) & = \int_\cC P^\mu \left(\frac{\nabla}{d\tau} \xi_\mu - \dot{Y}^\nu \xi_{\mu;\nu} \right) d\tau \\
& = \int_\cC P^\mu \frac{\nabla}{d\tau} \xi_\mu d\tau \\
& = - \int_\cC \xi_\mu \frac{\nabla}{d\tau} P^\mu d\tau
\end{align}
where the second line comes from $P^\mu \xi_{\mu;\nu} = 2 P^\mu \xi_\rho A^\rho{}_{\mu\nu} = 0$ since $P \propto l$. The third line arises after an integration by part, where the surface term vanishes since $\xi$ has compact support.

Since the above integral vanishes for all $\xi$ per the PGC, it requires
\begin{empheq}[box=\fbox]{equation}
\frac{\nabla P^\mu}{d\tau} = 0
\end{empheq}
This equation, together with $P \propto \dot{X} \propto l$, is the equation of motion for particles whose dipole and higher moments have been neglected. This coincides with (Finsler) geodesics on the manifold $M$.

Note that in the Finsler literature, other expressions for the Finsler geodesics are more common. Since $P \propto \dot{X}$, and $P^2 = \const$, we can write the above equation, setting $y^\mu = \dot{X}^\mu$, and using the definition of the spray coefficients $G^\mu$ \eqref{spray_coef}, as
\begin{equation}
\frac{d \dot{X}^\mu}{d\tau} = - G^\mu.
\end{equation}
which is a more usual expression for geodesics on Finsler manifolds, see \textit{e.g.} \cite{Rund59}.

Note that through this principle, we obtain in \eqref{solution_theta_geodesic} that $y \propto \dot{X}$, meaning that the curve on $TM_0$ is (up to a rescaling) the natural lift of the worldline on $M$. In a more general case, for example in the coming section that includes dipole effects, this result may not necessarily hold anymore.

\section{Distributions with dipole moment on Finsler space}
\label{s:FMPD}

If the particle to describe has internal structure in the form of a dipole moment, for instance angular momentum, the distribution \eqref{general_distrib} should contain terms up to a dipole contribution,
\begin{equation}
\cT(\delta g) = \half \int_\cC \left(\theta(\delta g) + \bar{\Phi}(\nabla \delta g) \right) d\tau,
\end{equation}
As we have seen in the previous section for geodesics, the variation $\delta g$ will be generated by the Lie derivative of the metric along some vector $\xi \in \Vect_c(M)$. The Lie derivative is purely horizontal which means that $\theta = \theta^{\mu\nu}(\bx, \by) \partial_{x^\mu} \otimes \partial_{x^\nu}$ is a symmetric twice contravariant tensor on $\pi^* TM$, however, due to the definition \eqref{cov_der} of the covariance derivative, $\nabla \delta g$ has a vertical component, which means that $\bar{\Phi}$ should have as well. We hence decompose $\bar{\Phi}$ into two terms, such that 
\begin{equation}
\label{def_omega}
\bar{\Phi}(\nabla \delta g) = \Phi^{\rho\mu\nu}  \delta g_{\mu\nu|\rho} + \Omega^{\rho\mu\nu} \delta g_{\mu\nu;\rho}
\end{equation}

In short, the distribution we need to consider is,
\begin{equation}
\label{space_distribution_dipole}
\cT(\delta g) = \half \int_\cC \left(\theta^{\mu\nu} \delta g_{\mu\nu} + \Phi^{\rho\mu\nu}  \delta g_{\mu\nu|\rho} + \Omega^{\rho\mu\nu} \delta g_{\mu\nu;\rho} \right) d\tau.
\end{equation}
Note that $\Phi^{\rho\mu\nu}$ and $\Omega^{\rho\mu\nu}$ are symmetric in their last two indices.

The distribution \eqref{space_distribution_dipole} involves the connection $\nabla$ explicitly, which we here choose to be Chern's connection. In Finsler geometry there is no unique choice of connection and a natural question is whether the distribution is invariant under a choice of connection. However, it is easy to see that the difference between Chern's connection and any other connection can be absorbed into a redefinition of $\theta$. Indeed, were one to use a connection $\widetilde{\omega}$ related to the Chern connection by $\widetilde{\omega}_\nu{}^\mu = \omega_\nu{}^\mu + D_\nu{}^\mu{}_{\rho} dx^\rho + E_\nu{}^\mu{}_{\rho} \frac{\delta y^\rho}{F}$, a redefinition of $\theta^{\mu\nu}$ as $\theta^{\mu\nu} - 2 \Phi^{\rho\lambda(\mu} D_\nu{}^{\mu)}{}_{\rho} - 2 \Omega^{\rho\lambda(\mu} E_\nu{}^{\mu)}{}_{\rho}$ would absorb the difference between connections. Hence, our distribution \eqref{space_distribution_dipole} is connection invariant.

The strategy to obtain the equations of motion is the same as in the previous section, in that we will obtain the expression of $\theta$ and $\Phi$ by considering specific transformations $\delta g$ such that they vanish on the worldline of the particle, but not their derivatives. We will first obtain $\Phi$ by considering a variation of the form $\delta g = L_{\alpha \beta \xi} g$ where both $\alpha$ and $\beta$ vanish on the curve, since the coefficients of $\Phi$ include second order derivatives. Then, transformations of the form $\delta g = L_{\alpha \xi} g$ will give us the form of $\theta$. The third, and ``vertical'', term $\Omega$ behaves differently in that we will deduce its expression from $\Phi$.

\subsection{The dipole moments}

Applying the the Principle of General Covariance \eqref{PGC} on the distribution \eqref{space_distribution_dipole} for a variation $\delta g = L_{\alpha \beta \xi} g$, with $\alpha|_C = \beta|_C = 0$, readily yields the condition,
\begin{equation}
\label{space_condition_dipole}
\partial_\mu \alpha \partial_\rho \beta \left(\Phi^{\rho \mu \nu} + \Phi^{\mu\rho\nu} + \Phi^{\rho \sigma \lambda} A_{\sigma\lambda}{}^\nu l^\mu + \Phi^{\mu \sigma \lambda} A_{\sigma\lambda}{}^\nu l^\rho\right) = 0
\end{equation}

This equation can be solved by expanding $\Phi$ in a series where terms depend on powers of $A$ and proceeding order by order. With some abuse of notation,
\begin{equation}
\Phi^{\rho\mu\nu} = \Phi_0^{\rho\mu\nu}(A = 0) + \Phi_1^{\rho\mu\nu}(A) + \Phi_2^{\rho\mu\nu}(A^2) + \ldots
\end{equation}

The first equation, obtained by collecting the terms not depending on $A$ is that of the Riemann case, whose solution is well known \cite{Souriau74},
\begin{equation}
\label{space_phi_riemann}
\Phi_0^{\rho\mu\nu} = \dot{X}^\rho B^{\mu\nu} + \half \left(S^{\rho\mu} \dot{X}^\nu + S^{\rho\nu} \dot{X}^\mu\right),
\end{equation}
for some symmetric 2-tensor $B$ and skewsymmetric 2-tensor $S$. 

The next order is linear in $A$, and the equation to solve is
\begin{equation}
\label{eq_t}
\partial_\mu \alpha \partial_\rho \beta \left( 
\Phi_1^{\rho\mu\nu} + \Phi_1^{\mu\rho\nu} + S^{\rho\sigma} \dot{X}^\lambda A_{\sigma\lambda}{}^\nu l^\mu + S^{\mu\sigma} \dot{X}^\lambda A_{\sigma\lambda}{}^\nu l^\rho
\right) = 0
\end{equation}

We can solve it, by being careful to remember the symmetry in $\mu, \nu$, as
\begin{equation}
\begin{split}
\Phi_1^{\rho\mu\nu} & = S^{(\mu\lambda} \dot{X}^\sigma l^{\nu)} A_{\lambda\sigma}{}^\rho - S^{\rho\lambda} \dot{X}^\sigma l^{(\mu} A_{\lambda\sigma}{}^{\nu)} - S^{(\mu\lambda} \dot{X}^\sigma l^\rho A_{\lambda\sigma}{}^{\nu)} \\
& \quad + \dot{X}^\rho B_1^{\mu\nu} + \half \left(S_1^{\rho\mu} \dot{X}^\nu + S_1^{\rho\nu} \dot{X}^\mu\right)
\end{split}
\end{equation}
where $B_1$ is an arbitrary symmetric 2-tensor depending linearly on $A$, and likewise $S_1$ an arbitrary skewsymmetric 2-tensor also depending linearly on $A$.

The equation quadratic in $A$ is then,
\begin{equation}
\begin{split}
\partial_\mu \alpha \partial_\rho \beta \big( 
\Phi_2^{\rho\mu\nu} + \Phi_2^{\mu\rho\nu} - 2 S^{\sigma\kappa} \dot{X}^\delta l^\rho A_{\kappa\delta}{}^\lambda A_{\sigma\lambda}{}^\nu l^\mu 
\big) = 0
\end{split}
\end{equation}
and the solution is
\begin{equation}
\begin{split}
\Phi_2^{\rho\mu\nu} & = 2 S^{\sigma\kappa} \dot{X}^\delta l^\rho l^{(\mu} A_{\sigma\lambda}{}^{\nu)} A_{\kappa\delta}{}^\lambda - S^{\sigma\kappa} \dot{X}^\delta l^\mu l^{\nu} A_{\kappa\delta}{}^{\lambda} A_{\sigma\lambda}{}^\rho \\
& \quad + \dot{X}^\rho B_2^{\mu\nu} + \half \left(S_2^{\rho\mu} \dot{X}^\nu + S_2^{\rho\nu} \dot{X}^\mu\right)
\end{split}
\end{equation}
where once again $B_2$ and $S_2$ are, respectively, symmetric and skewsymmetric and of quadratic order in $A$.

The next orders $n \ge 3$ are trivial, in the sense that they are all of the form $\Phi_n^{\rho\mu\nu} = \dot{X}^\rho B_n^{\mu\nu} + \half \left(S_n^{\rho\mu} \dot{X}^\nu + S_n^{\rho\nu} \dot{X}^\mu\right)$.

In the end, the solution for $\Phi$ is given by
\begin{equation}
\label{solution_phi}
\begin{split}
\Phi^{\rho\mu\nu} & = \dot{X}^\rho B^{\mu\nu} + \half \left(S^{\rho\mu} \dot{X}^\nu + S^{\rho\nu} \dot{X}^\mu\right) \\
& \quad + S^{(\mu\lambda} \dot{X}^\sigma l^{\nu)} A_{\lambda\sigma}{}^\rho - S^{\rho\lambda} \dot{X}^\sigma l^{(\mu} A_{\lambda\sigma}{}^{\nu)} - S^{(\mu\lambda} \dot{X}^\sigma l^\rho A_{\lambda\sigma}{}^{\nu)} \\
& \quad + 2 S^{\delta\lambda} \dot{X}^\kappa l^\rho l^{(\mu} A_{\gamma\lambda}{}^{\nu)} A_{\kappa\delta}{}^\gamma - S^{\delta\lambda} \dot{X}^\kappa l^\mu l^{\nu} A_{\gamma\lambda}{}^{\rho} A_{\kappa\delta}{}^\gamma
\end{split}
\end{equation}
where $B$ and $S$ are tensor that may depend arbitrarily on $A$.

For later convenience, we split the solution of $\Phi$ into its Riemann and Finsler parts,
\begin{equation}
\label{phi_riemann_T}
\Phi^{\rho\mu\nu} = \dot{X}^\rho B^{\mu\nu} + \half \left(S^{\rho\mu} \dot{X}^\nu + S^{\rho\nu} \dot{X}^\mu\right) + T^{\rho\mu\nu}
\end{equation}
where
\begin{equation}
\label{expr_t}
\begin{split}
T^{\rho\mu\nu} & = S^{(\mu\lambda} \dot{X}^\sigma l^{\nu)} A_{\lambda\sigma}{}^\rho - S^{\rho\lambda} \dot{X}^\sigma l^{(\mu} A_{\lambda\sigma}{}^{\nu)} - S^{(\mu\lambda} \dot{X}^\sigma l^\rho A_{\lambda\sigma}{}^{\nu)} \\
& \quad + 2 S^{\delta\lambda} \dot{X}^\kappa l^\rho l^{(\mu} A_{\gamma\lambda}{}^{\nu)} A_{\kappa\delta}{}^\gamma - S^{\delta\lambda} \dot{X}^\kappa l^\mu l^{\nu} A_{\gamma\lambda}{}^{\rho} A_{\kappa\delta}{}^\gamma
\end{split}
\end{equation}

Since none of the components of the tensor $T$ are directly parallel to $\dot{X}$, the linear and quadratic equations we solved should vanish regardless of the contraction with $\partial_\mu \alpha \partial_\rho \beta$. For future reference, this implies the skew symmetriness of the following expression in its $\rho$ and $\mu$ indices,
\begin{equation}
\label{T_expression_skew}
\begin{split}
T^{\rho \mu \nu} &  + l^{\mu} T^{\rho \sigma \lambda} A_{\sigma\lambda}{}^\nu + l^{\mu} S^{\rho \sigma} \dot{X}^\lambda A_{\sigma\lambda}{}^\nu = \\
& \left(A_{\sigma\lambda}{}^{[\rho} l^{\mu]} S^{\nu\lambda} - A_{\sigma\lambda}{}^\nu l^{[\rho} S^{\mu]\lambda} + A_{\sigma\lambda}{}^{[\rho} S^{\mu]\lambda} l^\nu - 2 A_{\sigma\lambda}{}^\kappa l^\nu S^{\delta\lambda} A_{\kappa\delta}{}^{[\rho} l^{\mu]} \right) \dot{X^\sigma}
\end{split}
\end{equation}

Putting the expression for $\Phi$ \eqref{phi_riemann_T} in the distribution \eqref{space_distribution_dipole} yields\footnote{For the sake of compactness, we will keep using the tensor $T$ instead of its expression \eqref{expr_t} unless where relevant.},
\begin{equation}
\cT(\delta g) = \half \int_\cC \left(\theta^{\mu\nu} \delta g_{\mu\nu} + \left(\dot{X}^\rho B^{\mu\nu} + S^{\rho\mu} \dot{X}^\nu + T^{\rho\mu\nu} \right) \delta g_{\mu\nu|\rho} + \Omega^{\rho\mu\nu} \delta g_{\mu\nu;\rho}\right)d\tau
\end{equation}
The ``$B$ term'' can be rewritten $\dot{X}^\rho B^{\mu\nu} \delta g_{\mu\nu|\rho} = B^{\mu\nu} \nabla \delta g_{\mu\nu}/d\tau - \dot{Y}^\rho B^{\mu\nu} \delta g_{\mu\nu;\rho}$. After an integration by part, this shows that this term can be absorbed into $\theta$ and $\Omega$ with the redefinitions $\theta^{\mu\nu}$ to $\theta^{\mu\nu} - \nabla B^{\mu\nu}/d\tau$ and $\Omega^{\rho\mu\nu}$ to $\Omega^{\rho\mu\nu} + \dot{Y}^\rho B^{\mu\nu}$.  Hence, the above integral is equivalent to,
\begin{equation}
\label{space_distribution_dipole2}
\cT(\delta g) = \half \int_\cC \left(\theta^{\mu\nu} \delta g_{\mu\nu} + S^{\rho\mu} \dot{X}^\nu \delta g_{\mu\nu|\rho} + T^{\rho\mu\nu} \delta g_{\mu\nu|\rho} + \Omega^{\rho\mu\nu} \delta g_{\mu\nu;\rho}\right)d\tau,
\end{equation}
meaning that $B$ in the definition \eqref{phi_riemann_T} of $\Phi$ does not actually contain additional degrees of freedom, just as in the Riemannian case \cite{Souriau74}.

Let us now focus on the second term in the above integral. On account of the skewsymmetry of $S$, 
\begin{equation}
S^{\rho\mu} \dot{X}^\nu \delta g_{\mu\nu|\rho} = \half S^{\rho\mu} \dot{X}^\nu \left(\delta g_{\mu\nu|\rho} - \delta g_{\rho\nu|\mu}\right)
\end{equation}
Hence we need to compute $\delta g_{\mu\nu|\rho} - \delta g_{\rho\nu|\mu}$ for $\delta g = L_{\bar{\xi}} g$. We find, using the formula for the commutation of derivatives \eqref{interchangehh} and Bianchi's first identity,
\begin{equation}
\begin{split}
L_{\bar{\xi}} g_{\mu\nu|\rho} - L_{\bar{\xi}} g_{\rho\nu|\mu} = & - 2 R_\nu{}^\lambda{}_{\rho\mu}\xi_\lambda + (\xi_{\mu|\rho} - \xi_{\rho|\mu})_{|\nu} + 2 l^\sigma \left[\left(A_{\mu\nu\lambda} \xi^\lambda{}_{|\sigma}\right){}_{|\rho} - \left(A_{\rho\nu\lambda} \xi^\lambda{}_{|\sigma}\right){}_{|\mu}\right] \\
& - 2 A_{\nu \lambda \kappa} \xi^\kappa R_{\sigma}{}^\lambda{}_{\rho\mu} l^\sigma - 2 A_{\mu \lambda \kappa} \xi^\kappa R_{\sigma}{}^\lambda{}_{\rho\nu} l^\sigma + 2 A_{\rho \lambda \kappa} \xi^\kappa R_{\sigma}{}^\lambda{}_{\mu\nu} l^\sigma
\end{split}
\end{equation}

We thus have,
\begin{equation}
\label{Slxigh}
\begin{split}
S^{\rho\mu} \dot{X}^\nu L_{\bar{\xi}} g_{\mu\nu|\rho} = S^{\rho\mu} \dot{X}^\nu \Big(& \xi_{\mu|\rho|\nu} - R_\nu{}^\lambda{}_{\rho\mu} \xi_\lambda - A_{\nu \lambda \kappa} \xi^\kappa R_{\sigma}{}^\lambda{}_{\rho\mu} l^\sigma + 2 A_{\rho \lambda \kappa} \xi^\kappa R_{\sigma}{}^\lambda{}_{\mu\nu} l^\sigma \\
& + 2 l^\sigma \left(A_{\mu\nu\lambda} \xi^\lambda{}_{|\sigma}\right){}_{|\rho} \Big)
\end{split}
\end{equation}

The first term of this relation is particularly promising, as it looks like we will be able to integrate it by parts to get $\nabla S/d\tau$. Since $\nabla \xi_{\mu|\rho} / d\tau = \dot{X}^\nu \xi_{\mu|\rho|\nu} + \dot{Y}^\nu \xi_{\mu|\rho;\nu}$, and remembering the interchange formula \eqref{interchangehv}, we find,
\begin{align}
\begin{split}
\int S^{\rho\mu} \dot{X}^\nu \xi_{\mu|\rho|\nu} d\tau & = - \int \left( \frac{\nabla S^{\rho\mu}}{d\tau} + 2 S^{\rho\sigma} \dot{Y}^\nu A^\mu{}_{\sigma\nu} \right) \xi_{\mu|\rho} d\tau  \\
& \quad - \int S^{\rho\mu} \dot{Y}^\nu \left(\cP_\mu{}^\lambda{}_{\rho\nu} - 2 A^\lambda{}_{\mu\sigma} \dot{A}^\sigma{}_{\rho\nu} + 2 A^\lambda{}_{\mu\nu|\rho}\right) \xi_\lambda d\tau
\end{split}
\\
\label{ipp_s}
& = \int \bigg[\dot{Y}^\nu \cP(S)^\lambda{}_\nu \xi_\lambda - \left( \frac{\nabla S^{\rho\mu}}{d\tau} + 2 S^{\rho\sigma} \dot{Y}^\nu A^\mu{}_{\sigma\nu} \right) \xi_{\mu|\rho} \bigg]d\tau
\end{align}
due to $\cP$ being symmetric in its first and third index, and the definition \eqref{ps}.

\smallskip

For the third term of the distribution \eqref{space_distribution_dipole2}, we will simply write it as
\begin{equation}
\label{Tlxigh}
T^{\rho\mu\nu} (L_{\bar{\xi}} g_{\mu\nu})_{|\rho} = 2 T^{\rho\mu\nu} \xi_{\mu|\nu|\rho} + 2 T^{\rho\mu\nu} l^\sigma A_{\mu\nu\lambda|\rho} \xi^\lambda{}_{|\sigma} + 2 T^{\rho\mu\nu} l^\sigma A_{\mu\nu\lambda} \xi^\lambda{}_{|\sigma|\rho}
\end{equation}

\smallskip

The fourth term $\Omega^{\rho\mu\nu} \delta g_{\mu\nu;\rho}$ requires some work. First, let us evaluate $(L_{\bar{\xi}} g_{\mu\nu})_{;\rho}$.
While $\xi^\lambda{}_{;\rho} = 0$, $\xi^\lambda{}_{|\mu;\rho}$ does not vanish, for we need to be careful about the commutative of derivatives \eqref{interchangehv} which shows,
\begin{equation}
\xi^\lambda{}_{|\mu;\rho} = - \xi^\sigma P_\sigma{}^\lambda{}_{\mu\rho}
\end{equation}

We then find, using the decomposition of $\cP$ in derivatives of $A$,
\begin{equation}
\label{lxigv}
\begin{split}
(L_{\bar{\xi}} g_{\mu\nu})_{;\rho} & = 2 \xi^\sigma A_{\mu\nu\rho|\sigma} + 2 \xi^\lambda{}_{|\sigma} \left(A_{\mu\nu\lambda;\rho} l^\sigma - A_{\mu\nu\lambda} l^\sigma l_\rho \right) \\
& \quad + 2 \left(A_{\lambda\nu\rho} \xi^\lambda{}_{|\mu} + A_{\lambda\mu\rho} \xi^\lambda{}_{|\nu} + A_{\mu\nu\lambda} \xi^\lambda{}_{|\rho}\right)
\end{split}
\end{equation}

Before reasoning on the expression of $\Omega$, let us remark that $H^{\rho(\mu} V^{\nu)} (L_{\bar{\xi}} g_{\mu\nu})_{;\rho} = 0$ for any skewsymmetric $H$ and any vector $V$, which is due to the symmetry properties of \eqref{lxigv} and the useful formula $A_{\rho\mu\nu;\lambda} - A_{\rho\mu\lambda;\nu} = A_{\rho\mu\nu} l_{\lambda} - A_{\rho\mu\lambda} l_{\nu}$ \cite{BaoCS00}.

Unlike for $\Phi$ and $\theta$, there seems to be no way to impose a constraint on $\Omega$ through the PGC. However, we should keep in mind that $\Omega$ is the vertical component to the dipole moment. It is reasonable to assume that the degrees of freedom of a dynamical system only stem from the purely horizontal terms, since these are the ones that can be measured physically, and since the vertical Finsler components are usually obtained through some compatibility relation, for example natural lifts. Hence, we assume that $\Omega$ cannot contain any new degree of freedom not present in $\Phi$. In short, $\Omega$ may only depend linearly on $S$ or its derivatives. Given the symmetry of $\Omega$, suitable terms are of the form $S^{\rho(\mu} \dot{Y}^{\nu)}$, $\frac{dS^{\rho(\mu}}{d\tau} \dot{X}^{\nu)}$, \ldots All these terms are of the form $H^{\rho(\mu} V^{\nu)}$ for skewsymmetric $H$, which implies that all $\Omega$ terms vanish from the distribution, per the previous paragraph,
\begin{equation}
\label{Omega_L}
\Omega^{\rho\mu\nu} (L_{\bar{\xi}} g_{\mu\nu})_{;\rho} = 0
\end{equation}
The above assumption will also be justified a posteriori in section \ref{s:spinoptics} where we will specialize our equations to recover a known model of Finsler spinoptics \cite{Duval08}, which does not contain any $\Omega$ term. Regardless, keeping track of $\Omega$ terms is not difficult, and we will still present the final equations of motion with these terms if one wishes for a different expression.

Reporting the results \eqref{Slxigh}, \eqref{ipp_s} \eqref{Tlxigh}, and \eqref{Omega_L} in the distribution \eqref{space_distribution_dipole2}, and using that $T$ is symmetric in its last 2 indices, we get,
\begin{equation}
\label{distribution_many_terms}
\begin{split}
\cT(L_{\bar{\xi}} g) = \int_\cC \bigg\lbrace & \half  \bigg[ S^{\rho\mu} \dot{X}^\nu \left(2 A_{\rho\kappa}{}^\lambda R_\sigma{}^\kappa{}_{\mu\nu}l^\sigma - R_\nu{}^\lambda{}_{\rho\mu} - A_{\nu\kappa}{}^\lambda R_\sigma{}^\kappa{}_{\rho\mu} l^\sigma\right) \\
& + \dot{Y}^\nu \cP(S)^\lambda{}_\nu \bigg] \, \xi_\lambda + \bigg[ \theta^{\mu\lambda} - \half \frac{\nabla S^{\mu\lambda}}{d\tau} + S^{\rho\sigma} \dot{X}^\nu l^\mu A_{\sigma\nu}{}^\lambda{}_{|\rho}  \\
&  + \theta^{\sigma\nu}A_{\sigma\nu}{}^\lambda l^\mu + T^{\rho\sigma\nu} l^\mu A_{\sigma\nu}{}^\lambda{}_{|\rho} - S^{\mu\sigma} \dot{Y}^\nu A^\lambda{}_{\sigma\nu} \bigg] \xi_{\lambda|\mu}  \\
& + \Big[ T^{\nu\mu\lambda} + T^{\nu\rho\sigma}A_{\rho\sigma}{}^\lambda l^\mu + S^{\nu\rho} \dot{X}^\sigma l^\mu A_{\rho\sigma}{}^\lambda \Big] \xi_{\lambda|\mu|\nu}\bigg\rbrace d\tau 
\end{split}
\end{equation}

Notice that the expression in brackets on the last line, which is contracted with $\xi_{\lambda|\mu|\nu}$, is exactly the expression that we found to be skew symmetric in some of its indices (here, $\nu$ and $\mu$) and that we calculated in \eqref{T_expression_skew}. This property, and the formula \eqref{interchangehh}, means that we can rewrite the last line of \eqref{distribution_many_terms} as,
\begin{align*}
& \Big( T^{\nu\mu\lambda} + T^{\nu\rho\sigma}A_{\rho\sigma}{}^\lambda l^\mu + S^{\nu\rho} \dot{X}^\sigma l^\mu A_{\rho\sigma}{}^\lambda\Big) \xi_{\lambda|\mu|\nu} \\
& \qquad = \half \left( T^{\nu\mu\lambda} + T^{\nu\rho\sigma}A_{\rho\sigma}{}^\lambda l^\mu + S^{\nu\rho} \dot{X}^\sigma l^\mu A_{\rho\sigma}{}^\lambda \right) \left(\xi_{\lambda|\mu|\nu} - \xi_{\lambda|\nu|\mu}\right) \\
& \qquad = \half \left( T^{\nu\mu\lambda} + T^{\nu\rho\sigma}A_{\rho\sigma}{}^\lambda l^\mu + S^{\nu\rho} \dot{X}^\sigma l^\mu A_{\rho\sigma}{}^\lambda \right) \left(R_\lambda{}^\kappa{}_{\mu\nu} + 2 A_{\lambda\beta}{}^\kappa l^\alpha R_\alpha{}^\beta{}_{\mu\nu} \right)\xi_\kappa
\end{align*}
The last term is due to $\xi_{\lambda;\kappa} = 2 \xi^\rho A_{\rho\lambda\kappa}$.

The integral then becomes (while taking the opportunity for some index relabeling),
\begin{align}
\label{distribution_many_terms_2}
\cT(L_{\bar{\xi}} g) = \int_\cC \bigg\lbrace & \half \bigg[  S^{\rho\mu} \dot{X}^\nu \left(2 A_{\rho\kappa}{}^\lambda R_\sigma{}^\kappa{}_{\mu\nu}l^\sigma - R_\nu{}^\lambda{}_{\rho\mu} - A_{\nu\kappa}{}^\lambda R_\sigma{}^\kappa{}_{\rho\mu} l^\sigma\right) \nonumber \\
& - \left( T^{\mu\rho\nu} + T^{\mu\kappa\sigma}A_{\kappa\sigma}{}^\rho l^\nu + S^{\mu\kappa} \dot{X}^\sigma l^\nu A_{\kappa\sigma}{}^\rho \right) \left(R_\rho{}^\lambda{}_{\mu\nu} + 2 A_{\rho\beta}{}^\lambda l^\alpha R_\alpha{}^\beta{}_{\mu\nu} \right) \nonumber \\
& + \dot{Y}^\mu \cP(S)^\lambda{}_\mu \bigg] \, \xi_\lambda  \\
& + \bigg[ \theta^{\mu\lambda} - \half \frac{\nabla S^{\mu\lambda}}{d\tau} + S^{\rho\sigma} \dot{X}^\nu l^\mu A_{\sigma\nu}{}^\lambda{}_{|\rho} + \theta^{\sigma\nu}A_{\sigma\nu}{}^\lambda l^\mu + T^{\rho\sigma\nu} l^\mu A_{\sigma\nu}{}^\lambda{}_{|\rho} \nonumber \\
& - S^{\mu\sigma} \dot{Y}^\nu A^\lambda{}_{\sigma\nu} \bigg] \xi_{\lambda|\mu} \bigg\rbrace d\tau \nonumber
\end{align}

Our distribution is now in the form where we have two groups of terms, one depending on $\xi$ and one depending on the first derivative of $\xi$.

\subsection{The monopole moment}

To obtain the expression of $\theta$, we will use the substitution $\xi \mapsto \alpha \xi$, where once again $\alpha|_\cC = 0$. The strategy is the same as earlier, but with one derivative instead of two. The distribution \eqref{distribution_many_terms_2} still vanishes, as per the Principle of General Covariance \eqref{PGC}, provided the expression in the bracket contracted with $\xi_{\lambda|\mu}$ vanishes when contracted with $\partial_\mu \alpha$ instead,
\begin{equation}
\begin{split}
\bigg[ & \theta^{\mu\lambda} - \half \frac{\nabla S^{\mu\lambda}}{d\tau} + S^{\rho\sigma} \dot{X}^\nu l^\mu A_{\sigma\nu}{}^\lambda{}_{|\rho} + \theta^{\sigma\nu}A_{\sigma\nu}{}^\lambda l^\mu \\
& + T^{\rho\sigma\nu} l^\mu A_{\sigma\nu}{}^\lambda{}_{|\rho} - S^{\mu\sigma} \dot{Y}^\nu A^\lambda{}_{\sigma\nu} \bigg] \partial_\mu \alpha = 0,
\end{split}
\end{equation}

This is solved in a similar way as for the derivation of the geodesics in the previous section, \textit{i.e.} we must have
\begin{equation}
\label{condition_pgc_1}
\begin{split}
& \theta^{\mu\lambda} - \half \frac{\nabla S^{\mu\lambda}}{d\tau} + S^{\rho\sigma} \dot{X}^\nu l^\mu A_{\sigma\nu}{}^\lambda{}_{|\rho} + \theta^{\sigma\nu}A_{\sigma\nu}{}^\lambda l^\mu \\
& + T^{\rho\sigma\nu} l^\mu A_{\sigma\nu}{}^\lambda{}_{|\rho} - S^{\mu\sigma} \dot{Y}^\nu A^\lambda{}_{\sigma\nu} = \dot{X}^\mu P^\lambda,
\end{split}
\end{equation}
for some tensor $P$, yet to be determined.

Given the symmetry of $\theta$ and the skewsymmetry of $\dot{S}$, this is readily solved as,
\begin{subequations}
\begin{align}
\begin{split}
\theta^{\mu\lambda} & = P^{(\mu} \dot{X}^{\lambda)} - P^\sigma \dot{X}^\nu l^{(\mu} A_{\sigma\nu}{}^{\lambda)} - S^{\rho\sigma} \dot{X}^\nu l^{(\mu} A_{\sigma\nu}{}^{\lambda)}{}_{|\rho} - T^{\rho\sigma\nu} l^{(\mu} A_{\sigma\nu}{}^{\lambda)}{}_{|\rho} \\
& \quad + A^{(\lambda}{}_{\sigma\nu} S^{\mu)\sigma} \dot{Y}^\nu + \half l^{(\mu} \widehat{Q}(S) ^{\lambda)}{}_\nu  \dot{Y}^\nu
\end{split}
\\
\label{sol_sdot}
\begin{split}
\half \frac{\nabla S^{\mu\lambda}}{d\tau} & = P^{[\mu} \dot{X}^{\lambda]} + P^\sigma \dot{X}^\nu l^{[\mu} A_{\sigma\nu}{}^{\lambda]} + S^{\rho\sigma} \dot{X}^\nu l^{[\mu} A_{\sigma\nu}{}^{\lambda]}{}_{|\rho} + T^{\rho\sigma\nu} l^{[\mu} A_{\sigma\nu}{}^{\lambda]}{}_{|\rho} \\
& \quad - A^{[\lambda}{}_{\sigma\nu} S^{\mu]\sigma} \dot{Y}^\nu - \half l^{[\mu} \widehat{Q}(S) ^{\lambda]}{}_\nu  \dot{Y}^\nu
\end{split}
\end{align}
\end{subequations}

We know now that the whole expression in bracket in front of $\xi_{\lambda|\mu}$ in \eqref{distribution_many_terms_2} reduces to $\dot{X}^\mu P^\lambda$. The distribution is now
\begin{equation*}
\begin{split}
\cT(L_{\bar{\xi}} g) & = \int_\cC \bigg[\dot{X}^\mu P^\lambda \xi_{\lambda|\mu} - \half S^{\rho\mu} \dot{X}^\nu \left(R_\nu{}^\lambda{}_{\rho\mu} + A_{\nu\kappa}{}^\lambda R_\sigma{}^\kappa{}_{\rho\mu} l^\sigma - 2 A_{\rho\kappa}{}^\lambda R_\sigma{}^\kappa{}_{\mu\nu}l^\sigma \right) \xi_\lambda \\
& \qquad - \half \left( S^{\mu\kappa} \dot{X}^\sigma l^\nu A_{\kappa\sigma}{}^\rho + T^{\mu\rho\nu} + T^{\mu\kappa\sigma}A_{\kappa\sigma}{}^\rho l^\nu\right) \left(R_\rho{}^\lambda{}_{\mu\nu} + 2 A_{\rho\beta}{}^\lambda l^\alpha R_{\alpha}{}^\beta{}_{\mu\nu} \right) \xi_\lambda \\
& \qquad + \half \dot{Y}^\mu \cP(S)^\lambda{}_\mu \xi_\lambda \bigg] d\tau 
\end{split}
\end{equation*}

An integration by parts to move the derivative from $\xi$ to $P$ yields
\begin{equation}
\label{ipp_p}
\int \dot{X}^\mu P^\lambda \xi_{\lambda|\mu} d\tau = - \int \frac{\nabla P^\lambda}{d\tau} \xi_\lambda d\tau - 2 \int P^\sigma \dot{Y}^\mu A^\lambda{}_{\sigma\mu} \xi_\lambda d\tau
\end{equation}
Note that now, unlike in the geodesic case, we do not have $P$ parallel to $l$ and $\dot{X}$.

Then, using \eqref{T_expression_skew}, and the Bianchi identity, we find the final form of the distribution,
\begin{equation*}
\label{final_distribution}
\begin{split}
\cT(L_{\bar{\xi}} g) & = - \int_\cC \bigg[ \frac{\nabla P^\lambda}{d\tau} + \half  R_\nu{}^\lambda{}_{\rho\mu} S^{\rho\mu} \dot{X}^\nu + \half S^{\rho\mu} \dot{X}^\nu A_{\nu\kappa}{}^\lambda R_{\sigma}{}^\kappa{}_{\rho\mu} l^\sigma - S^{\rho\mu} \dot{X}^\nu A_{\rho\kappa}{}^\lambda R_\sigma{}^\kappa{}_{\mu\nu} l^\sigma  \\
& \qquad - 2 S^{\nu\kappa} \dot{X}^\sigma l^{[\mu} A^{\rho]}{}_{\sigma\kappa} R_\rho{}^\lambda{} _{\mu\nu} + \half R^\lambda{}_\mu \widehat{Q}(S)^\mu{}_\sigma \dot{X}^\sigma - 2 S^{\kappa(\rho} A^{\nu)}{}_{\sigma\kappa} A_{\rho\mu}{}^\lambda R^\mu{}_\nu \dot{X}^\sigma \\
& \qquad  + 2 P^\sigma \dot{Y}^\mu A^\lambda{}_{\sigma\mu} - \half \dot{Y}^\mu \cP(S)^\lambda{}_\mu \bigg] \xi_\lambda d\tau
\end{split}
\end{equation*}

\subsection{Equations of motion}

The Principle of General Covariance states that the above distribution should vanish for all $\xi$ (with compact support). This clearly yields the equation,
\begin{equation}
\label{Pdot_unfinished}
\begin{split}
\frac{\nabla P^\lambda}{d\tau} = & - \half  R_\nu{}^\lambda{}_{\rho\mu} S^{\rho\mu} \dot{X}^\nu - \half S^{\rho\mu} \dot{X}^\nu A_{\nu\kappa}{}^\lambda R_{\sigma}{}^\kappa{}_{\rho\mu} l^\sigma + S^{\rho\mu} \dot{X}^\nu A_{\rho\kappa}{}^\lambda R_\sigma{}^\kappa{}_{\mu\nu} l^\sigma  \\
& + 2 S^{\nu\kappa} \dot{X}^\sigma l^{[\mu} A^{\rho]}{}_{\sigma\kappa} R_\rho{}^\lambda{} _{\mu\nu} - \half R^\lambda{}_\mu \widehat{Q}(S)^\mu{}_\sigma \dot{X}^\sigma + 2 S^{\kappa(\rho} A^{\nu)}{}_{\sigma\kappa} A_{\rho\mu}{}^\lambda R^\mu{}_\nu \dot{X}^\sigma \\
& - 2 P^\sigma \dot{Y}^\mu A^\lambda{}_{\sigma\mu}  + \half \dot{Y}^\mu \cP(S)^\lambda{}_\mu
\end{split}
\end{equation}

Some cosmetic work is in order. 
The Riemann tensor in Finsler geometry does not enjoy the skewsymmetriness property on its first two indices, as one has \cite{BaoCS00}
\begin{equation}
R_{\mu\nu\lambda\sigma} + R_{\nu\mu\lambda\sigma} = 2 B_{\mu\nu\lambda\sigma} = - 2 A_{\mu\nu\kappa} R_\rho{}^\kappa{}_{\lambda\sigma} l^\rho
\end{equation}
This implies on the exchange of pair of indices,
\begin{equation}
\label{diff_R_sym}
R_{\mu\nu\lambda\sigma} - R_{\lambda\sigma\mu\nu} = B_{\mu\nu\lambda\sigma} - B_{\lambda\sigma\mu\nu} + B_{\mu\sigma\nu\lambda} + B_{\nu\lambda\mu\sigma} + B_{\sigma\nu\lambda\mu} + B_{\lambda\mu\sigma\nu}
\end{equation}

This means that we need to be careful, since we defined in section \ref{s:finsler} $R(S)_{\mu\nu} = R_{\lambda\sigma\mu\nu} S^{\lambda\sigma} \neq R_{\mu\nu\lambda\sigma}S^{\lambda\sigma}$, the latter term being the one that appears in \eqref{Pdot_unfinished}. We can use the relation \eqref{diff_R_sym} to show,
\begin{equation}
\begin{split}
- \half R_{\nu\lambda\rho\mu} S^{\rho\mu} & =
- \half R(S)_{\nu\lambda}
+ \half A_{\nu\lambda}{}^\kappa R_{\sigma\kappa\rho\mu} l^\sigma S^{\rho\mu} \\
& \quad + A_{\nu\mu}{}^\kappa R_{\sigma\kappa\lambda\rho} l^\sigma S^{\rho\mu} 
- A_{\lambda\mu}{}^\kappa R_{\sigma\kappa\nu\rho} l^\sigma S^{\rho\mu} 
\end{split}
\end{equation}

Using the above relation and the Bianchi identity, it turns out that $\nabla P/d\tau$ considerably simplifies from \eqref{Pdot_unfinished} to,
\begin{equation}
\begin{split}
\frac{\nabla P^\lambda}{d\tau} & = \half R(S)^\lambda{}_\nu \dot{X}^\nu + \half \dot{Y}^\mu \cP(S)^\lambda{}_\mu  - 2 P^\sigma \dot{Y}^\mu A^\lambda{}_{\sigma\mu} 
\end{split}
\end{equation}

On the other hand, inserting the expression \eqref{expr_t} of $T$ into $\nabla S/d\tau$ \eqref{sol_sdot} yields,
\begin{equation}
\begin{split}
\frac{\nabla S^{\mu\lambda}}{d\tau} & = 2 P^{[\mu} \dot{X}^{\lambda]} + 2  \dot{X}^\nu  A^{[\lambda}{}_{\sigma\nu}{}l^{\mu]} P^\sigma + \dot{X}^\nu \cP(S)_\nu{}^{[\mu} l^{\lambda]} \\
& \quad - 2 \dot{Y}^\nu A^{[\lambda}{}_{\sigma\nu} S^{\mu]\sigma} -  \dot{Y}^\nu  \widehat{Q}(S)_\nu{}^{[\mu} l^{\lambda]}
\end{split}
\end{equation}

In conclusion, we find that the distribution of a dynamical system on its worldline in Finsler space is characterized by the quantities $(X, P, S)$ which obey the equations,
\begin{subequations}
\label{FinslerMPD_chern}
\begin{empheq}[box=\fbox]{align}
\begin{split}
\frac{\nabla P^\lambda}{d\tau} & = \half R(S)^\lambda{}_\nu \dot{X}^\nu + \half  \cP(S)^\lambda{}_\mu \dot{Y}^\mu  - 2 P^\sigma  A^\lambda{}_{\sigma\mu} \dot{Y}^\mu 
\end{split}
\\
\begin{split}
\label{fmpd_sdot}
\frac{\nabla S^{\mu\lambda}}{d\tau} & = 2 P^{[\mu} \dot{X}^{\lambda]} + 2  \dot{X}^\nu  A^{[\lambda}{}_{\sigma\nu}{}l^{\mu]} P^\sigma + \dot{X}^\nu \cP(S)_\nu{}^{[\mu} l^{\lambda]} \\
& \quad - 2 \dot{Y}^\nu A^{[\lambda}{}_{\sigma\nu} S^{\mu]\sigma} -  \dot{Y}^\nu  \widehat{Q}(S)_\nu{}^{[\mu} l^{\lambda]}
\end{split}
\end{empheq}
\end{subequations}
These equations reduce to the MPD equations in the Riemannian case, \textit{i.e.} when $A = 0$. The similar form to the MPD equations prompts us to call these equations the Finsler-MPD (FMPD) equations.

The directional derivatives in the equations \eqref{FinslerMPD_chern} are written using the Chern connection. We can use another connection, for example that of Cartan. Since the connections are related by $\widehat{\omega}_\nu{}^\mu = \omega_\nu{}^\mu + A^\mu{}_{\nu\lambda} \frac{\delta y^\lambda}{F}$, we find,
\begin{subequations}
\label{FinslerMPD_cartan}
\begin{empheq}[box=\fbox]{align}
\frac{\widehat{\nabla} P^\lambda}{d\tau} & = \half R(S)^\lambda{}_\nu \dot{X}^\nu + \half  \cP(S)^\lambda{}_\mu \dot{Y}^\mu  - P^\sigma  A^\lambda{}_{\sigma\mu} \dot{Y}^\mu \\
\label{fmpd_sdot_cartan}
\frac{\widehat{\nabla} S^{\mu\lambda}}{d\tau} & = 2 P^{[\mu} \dot{X}^{\lambda]} + 2  \dot{X}^\nu  A^{[\lambda}{}_{\sigma\nu}{}l^{\mu]} P^\sigma + \dot{X}^\nu \cP(S)_\nu{}^{[\mu} l^{\lambda]} -  \dot{Y}^\nu  \widehat{Q}(S)_\nu{}^{[\mu} l^{\lambda]}
\end{empheq}
\end{subequations}
In the rest of this article, we will use the above equations of motion with the Cartan connection. The equations appear simpler and at the same time are easier to work with, as the Cartan connection is $g$-compatible which means we can freely lower and raise the indices, for example to consider $\widehat{\nabla} S^{\mu}{}_\lambda/d\tau$, without introducing new terms.

Now, $\dot{Y}$ can be written in terms of derivative of $l$, which will be useful for us later on. Indeed, consider $\widehat{\nabla} l^\mu/d\tau = \dot{X}^\rho l^\mu{}_{|\rho} + \dot{Y}^\rho l^\mu{}_{\hat{;}\rho} = \dot{Y}^\rho l^\mu{}_{;\rho} = \dot{Y}^\rho \left(\delta^\mu_\rho - l^\mu l_\rho\right)$. Hence,
\begin{equation}
\label{ydot_ldot}
\dot{Y}^\rho = \frac{\hat{\nabla} l^\rho}{d\tau} + l^\rho \dot{Y}^\sigma l_\sigma
\end{equation}

Conveniently, in the FMPD equations, $\dot{Y}$ is always contracted with an object that vanishes when contracted with $l$, meaning that we can readily replace $\dot{Y}$ by $\hat{\nabla} l/d\tau$ in the equations.

Note that, the FMPD equations share the properties of the MPD equations in (pseudo)-Riemannian spaces, in that there are not enough equations for all the parameters $(\dot{X}, \dot{Y}, P, S)$ describing the worldline, and one needs \emph{supplementary conditions} to close the system of equations.

There remains the question of interpretation of the variables that appear in these equations, namely $P$ and $S$. Just as for the Riemannian MPD equations, it seems natural to use these equations to describe dynamical systems where $P$ is the momentum of the system, and $S$ is its dipole moment, such as its angular momentum.

It is interesting that in the geodesic case, the condition \eqref{particle_pgc_condition} required that $P \propto \dot{X} \propto l$, but here the presence of $\nabla S/d\tau$ in the corresponding condition \eqref{condition_pgc_1} leaves $P$ without such a constraint. In other words, in the geodesics case, the momentum is necessarily along the direction of $l$ on the curve (or conversely), but when the dynamical system has internal degree of freedom in the form of $S$, its momentum on the curve is not necessarily parallel to $l$, nor to the velocity of the curve. That the momentum is not necessarily parallel to the velocity already happens on Riemannian geometries in similar situations. 

Until now, we have worked with objects homogeneous of degree 0. While it is convenient, this means that we cannot apply the equations \eqref{FinslerMPD_cartan} in some situations, notably to photons later on in this article. Thankfully, it is straightforward to transform all our objects back to inhomogeneous ones, \textit{e.g.} using $y$ and $C$ instead of $l$ and $A$ respectively. We have, using the simpler $\widehat{\nabla} y^\mu/d\tau = F \, \dot{Y}^\mu$, the inhomogeneous FMPD equations,
\begin{subequations}
\label{FinslerMPD_cartan_inhomogeneous}
\begin{empheq}[box=\fbox]{align}
\frac{\widehat{\nabla} P^\lambda}{d\tau} & = \half R(S)^\lambda{}_\nu \dot{X}^\nu + \half  \cP_I(S){}^\lambda{}_\mu \frac{\widehat{\nabla} y^\mu}{d\tau}  - P^\sigma  C^\lambda{}_{\sigma\mu} \frac{\widehat{\nabla} y^\mu}{d\tau} \\
\frac{\widehat{\nabla} S^{\mu\lambda}}{d\tau} & = 2 P^{[\mu} \dot{X}^{\lambda]} + 2  \dot{X}^\nu  C^{[\lambda}{}_{\sigma\nu} y^{\mu]} P^\sigma + \dot{X}^\nu \cP_I(S)_\nu{}^{[\mu} y^{\lambda]} -  \frac{\widehat{\nabla} y^\nu}{d\tau} \widehat{Q}_I(S)_\nu{}^{[\mu} y^{\lambda]}
\end{empheq}
\end{subequations}
where $\cP_I(S)$ and $\widehat{Q}_I(S)$ are the inhomogeneous versions of $\cP(S)$ and $\widehat{Q}(S)$, \textit{i.e.} $\cP_I(S)_{\mu\nu} = 2 \left(C_{\mu\nu\lambda|\sigma} - C_{\lambda\mu\kappa} C^\kappa{}_{\sigma\nu|\rho} y^\rho\right)S^{\lambda\sigma}$ and $\widehat{Q}_I(S)_{\mu\nu} = - 2 C_{\lambda\kappa\mu} C^\kappa{}_{\sigma\nu} S^{\lambda \sigma}$.

As a final comment in this section, had we chosen to keep a general $\Omega$ term earlier on, the equations of motion would be,
\begin{subequations}
\begin{empheq}{align}
\begin{split}
\frac{\nabla P^\lambda}{d\tau} & = \half R(S)^\lambda{}_\nu \dot{X}^\nu + \half  \cP(S)^\lambda{}_\mu \dot{Y}^\mu  - 2 P^\sigma  A^\lambda{}_{\sigma\mu} \dot{Y}^\mu  + 2\Omega^{\rho\mu\nu} A_{\rho\mu\nu|\kappa} g^{\kappa\lambda}
\end{split}
\\
\begin{split}
\frac{\nabla S^{\mu\lambda}}{d\tau} & = 2 P^{[\mu} \dot{X}^{\lambda]} + 2  \dot{X}^\nu  A^{[\lambda}{}_{\sigma\nu}{}l^{\mu]} P^\sigma + \dot{X}^\nu \cP(S)_\nu{}^{[\mu} l^{\lambda]} \\
& \quad - 2 \dot{Y}^\nu A^{[\lambda}{}_{\sigma\nu} S^{\mu]\sigma} -  \dot{Y}^\nu  \widehat{Q}(S)_\nu{}^{[\mu} l^{\lambda]} + \Omega^{\rho\sigma\nu} \left(l^{[\mu} A_{\sigma\nu}{}^{\lambda]}{}_{;\rho} - l^{[\mu} A_{\sigma\nu}{}^{\lambda]} l_\rho \right) \\
& \quad - \left(2 \Omega^{\rho\nu\kappa} - \Omega^{\nu\rho\kappa}\right) A_{\rho\kappa}{}^\sigma A_{\sigma\nu}{}^{[\lambda} l^{\mu]} + 2 \Omega^{\rho\nu[\mu} A_{\rho\nu}{}^{\lambda]} + A_{\rho\nu}{}^{[\lambda} \Omega^{\mu]\rho\nu}
\end{split}
\end{empheq}
\end{subequations}

\subsection{Conserved quantities}
\label{ss:cons_quant}

The conserved quantities depend on the symmetries of the equations of motion. When using a Lagrangian to describe the dynamical system, these quantities are obtained from the Noether theorem. Here of course, we do not have a Lagrangian and the system of equations is not closed. But the equations \eqref{FinslerMPD_chern} or \eqref{FinslerMPD_cartan} are still sufficient to compute the conserved quantities, just like the Riemannian case \cite{Souriau74}.

The distribution \eqref{space_distribution_dipole} can be put in the form
\begin{equation}
\cT(\delta g) = \int \text{surface} \, + \int \text{e.o.m} \, d\tau
\end{equation}
where $\int \text{surface}$ vanishes for the diffeomorphisms with compact support that we considered in the previous section, and e.o.m are the equations of motion \eqref{FinslerMPD_chern}.

Now, consider $\delta g = L_Z g = 0$, \textit{i.e.} a transformation generated by a Killing vector field. By definition, $\delta g = 0$, and it is immediate that we also have, on shell,
\begin{equation}
\cT(L_Z g) = 0
\end{equation}

The difference with the previous section is that we know the expression of $\cT(\delta g)$, and that Killing fields do not have compact support. The latter means that the surface terms do not vanish because of $\xi$ being with compact sypport and hence that the surface term should vanish for all $\xi$. Keeping track of the surface terms in the previous section, which appeared when we did integrations by parts to get $\nabla S/d\tau$ and $\nabla P/d\tau$, we find,
\begin{equation}
\cT(L_Z g) = \half \int \frac{\nabla}{d\tau} \left(S^{\mu\lambda} Z_{\lambda|\mu} \right) d\tau + \int \frac{\nabla}{d\tau} \left(P^\lambda Z_\lambda\right) d\tau
\end{equation}

Since $\cT(L_Z g) = 0$ for all Killing vectors $Z$, we have
\begin{equation}
\frac{\nabla}{d\tau} \Psi(Z) = 0
\end{equation}
where
\begin{equation}
\label{cons_quant}
\Psi(Z) = P^\lambda Z_\lambda + \half S^{\mu\lambda} Z_{\lambda|\mu}
\end{equation}
is conserved on the worldline, for $Z$ a Finsler Killing field. 

Since the surface terms do not depend on the Cartan tensor, the conserved quantities have the same form as for the standard MPD equations, see \textit{e.g.} \cite[\S 7]{Souriau74}.

This can also be directly, though tediously, verified by checking the vanishing of the derivative of $\Psi(Z)$ along the worldline,
\begin{align}
\frac{\nabla}{d\tau} \Psi(Z) = \frac{\nabla P^\lambda}{d\tau} Z_\lambda + P^\lambda \frac{\nabla Z_\lambda}{d\tau} + \half \frac{\nabla S^{\mu\lambda}}{d\tau} Z_{\lambda|\mu} + \half S^{\mu\lambda} \frac{\nabla Z_{\lambda|\mu}}{d\tau}
\end{align}
on account of the equations of motion \eqref{FinslerMPD_chern}, the definition of the derivative along the curve \eqref{def_cov_der_curve}, the (a)symmetry  of $R$ \eqref{diff_R_sym}, the Bianchi identity for $R$, the interchange formulas \eqref{interchangehh} and \eqref{interchangehv}, and the Killing equation for $g$ and $Z$ following \eqref{lie_der}.

\section{Supplementary conditions}
\label{s:supp_cond}

As explained above, the FMPD equations are not closed: we are missing equations for $\dot{X}$ (and $\dot{Y}$), just like with the standard MPD equations. One needs to posit \emph{supplementary conditions} to close the system of equations. 

The main difference with the Riemann MPD equations is that here the behavior of a dynamical system in a Finsler geometry depends both on its position and the ``direction'' it is facing. The first question is then what ``direction'' of the system defines the Finsler tangent direction $y$? There are two obvious candidates: the momentum $P$ and the velocity $\dot{X}$. While geodesic motion implies that the two are parallel, when considering dipole moments they define different directions in general, as is well known when working already on the MPD equation in pseudo-Riemannian geometry, see \textit{e.g.} \cite{CostaN14}, and as we will see later. The difference $\dot{X} - P$ is usually called \emph{anomalous velocity}. In short we have a choice:
\begin{equation}
P = y \quad \text{or} \quad \dot{X} = y \; ?
\end{equation}
Or perhaps a linear combination. In the rest of this article, we will assume the first choice, that it is the system's momentum $P$ that couples to the dynamics. Note that if one has $P^2 = m^2$, $P = y$ can equivalently be written $P = m l$. Our justification is that it is the momentum, and not the velocity, which defines a state of a dynamical system in phase space, and it is the momentum that usually couples to external fields. Moreover, this choice is also motivated in that it happens to neatly simplify the FMPD equations \eqref{FinslerMPD_cartan} to,
\begin{subequations}
\label{FinslerMPD_cartan_py}
\begin{empheq}{align}
\frac{\widehat{\nabla} P^\lambda}{d\tau} & = \half R(S)^\lambda{}_\nu \dot{X}^\nu + \half  \cP(S)^\lambda{}_\mu \dot{Y}^\mu \\
\label{fmpd_sdot_cartan_py}
\frac{\widehat{\nabla} S^{\mu\lambda}}{d\tau} & = 2 P^{[\mu} \dot{X}^{\lambda]} + \dot{X}^\nu \cP(S)_\nu{}^{[\mu} l^{\lambda]} -  \dot{Y}^\nu  \widehat{Q}(S)_\nu{}^{[\mu} l^{\lambda]}
\end{empheq}
\end{subequations}

Then, similarly to the usual Riemannian MPD equations, it is also necessary to posit some relation for the dipole moment $S$. In short, it is necessary to choose a reference frame to define the dipole moment \cite{CostaN14}. There are different choices, depending on how one wishes to define the worldline, but two main choices are the Pirani condition \cite{Pirani56},
\begin{equation}
S^\mu{}_\nu \dot{X}^\nu = 0
\end{equation}
the Tulczyjew condition \cite{Tulczyjew59}
\begin{equation}
S^\mu{}_\nu P^\nu = 0
\end{equation}
and the Corinaldesi-Papapetrou condition \cite{CorinaldesiP51},
\begin{equation}
S^\mu{}_\nu t^\nu = 0
\end{equation}
where $t$ is the vector field representing the velocity of the observer measuring the system.

Note that the kernel of $S$ is necessarily even, since $S$ is skewsymmetric. This means that in even dimensions, $\ker(S)$ is generated by at least 2 directions, and hence that several of these conditions can hold at the same time. 

In the rest of the paper, we will assume the Tulczyjew condition $S^\mu{}_\nu P^\nu = 0$, together with the Corinaldesi-Papapetrou condition $S^\mu{}_\nu t^\nu = 0$ in dimension 4.

\bigskip

It remains to investigate the compatibility of these supplementary conditions with the equations of motion \eqref{FinslerMPD_cartan_py}.

(Elementary) dynamical systems are classified by invariant numbers, called Casimir invariants. These quantities are always conserved, as a consequence of the full group of symmetry of the underlying geometry\footnote{In particular, such Casimir invariants classify the different group representations of the symmetry group.}. For example on 4 dimensional spacetime in Physics, it is common to call the two Casimir invariants the mass $m$ and the (longitudinal, or scalar) spin $s$. 

\bigskip

Here, we would like to define 
\begin{subequations}
\begin{align}
p^2 & = P^\mu P_\mu \\
s^2 & = \frac{1}{2} S^{\mu\nu} S_{\mu\nu}
\end{align}
\end{subequations}

However for these definitions to make sense, we need to prove that these quantities are always constant along the motion.

For the scalar spin, we are interested in computing $\frac{\widehat{\nabla} S^{\mu\nu}}{d\tau} S_{\mu\nu}$. From the equation of motion \eqref{fmpd_sdot_cartan}, we immediately notice that the supplementary conditions $S^\mu{}_\nu P^\nu = 0$ and $P \propto l$ imply $\frac{\widehat{\nabla} S^{\mu\nu}}{d\tau} S_{\mu\nu} = 0$ and hence that the scalar spin is conserved.

Then, we want to prove that $P^2 = \const$. From $S^\mu{}_\nu P^\nu = 0$ we deduce that $\frac{\widehat{\nabla} S^\mu{}_\nu}{d\tau} P^\nu + S^\mu{}_\nu \frac{\widehat{\nabla} P^\nu}{d\tau} = 0$, which means that, by skewsymmetry of $S$,
\begin{align}
0 & =  \frac{\widehat{\nabla} P_\mu}{d\tau} \frac{\widehat{\nabla} S^{\mu\nu}}{d\tau} P_\nu \\
& = p^2 \frac{\widehat{\nabla} P_\nu}{d\tau} \dot{X}^\nu - P_\nu \dot{X}^\nu P_\mu \frac{\widehat{\nabla} P^\mu}{d\tau} - \frac{p}{2} \dot{X}^\nu \cP(S)_\nu{}^\lambda \frac{\widehat{\nabla} P_\lambda}{d\tau} \\
& = - P_\nu \dot{X}^\nu P_\mu \frac{\widehat{\nabla} P^\mu}{d\tau}
\end{align}
This leaves two possibilities, either $P_\mu \frac{\widehat{\nabla} P^\mu}{d\tau} = 0$ and we are finished proving that $P^2 = \const$, or
\begin{equation}
P_\nu \dot{X}^\nu = 0
\end{equation}

\section{Finsler space of dimension 3: Finsler spinoptics}
\label{s:spinoptics}

Specific examples of dynamical systems with dipole moments in Finsler geometry have already been studied in the literature, although not in a systematic way. For example Finsler spinoptics, where one considers the motion of light, including polarization effects, in a 3 dimensional manifold, has been worked out in \cite{Duval08}. Such a study brings us the opportunity to recover their equations, obtained in an independent way through symplectic geometry, from our FMPD equations, as a check and as a test of supplementary conditions.

It is clear from \cite{Duval08} that the following relations hold, which will serve as our supplementary conditions,
\begin{subequations}
\label{chd_suppl}
\begin{align}
P^\mu & = p l^\mu  \label{chd_def_p} \\
S^\mu{}_\nu P^\nu & = 0 \label{chd_tul}
\end{align}
\end{subequations}
For some scalar $p$ describing the ``color'' (wavelength) of light. Note that the Tulczyjew condition in 3 dimensions is equivalent to express the dipole moment tensor as
\begin{equation}
\label{chd_def_s}
S_{\mu\nu} = s \epsilon_{\mu\nu\lambda} l^\lambda
\end{equation}
where $s$ is the ``scalar spin'' of the dynamical system. In particular, its sign will describe the polarization, or helicity, of light.

\bigskip

Using \eqref{chd_def_p} and \eqref{ydot_ldot}, we have $\dot{Y}^\mu = \frac{1}{p} \frac{\widehat{\nabla} P^\mu}{d\tau}$, and hence the FMPD equations are,
\begin{subequations}
\begin{align}
\frac{\widehat{\nabla} P^\lambda}{d\tau} & = \half R(S)^\lambda{}_\nu \dot{X}^\nu + \frac{1}{2p}  \cP(S)^\lambda{}_\mu \frac{\widehat{\nabla} P^\mu}{d\tau} \\
\frac{\widehat{\nabla} S^{\mu\lambda}}{d\tau} & = 2 P^{[\mu} \dot{X}^{\lambda]} + \dot{X}^\nu \cP(S)_\nu{}^{[\mu} l^{\lambda]} - \frac{1}{p} \frac{\widehat{\nabla} P^\nu}{d\tau} \widehat{Q}(S)_\nu{}^{[\mu} l^{\lambda]} \label{sdot_pydot}
\end{align}
\end{subequations}

Since the following is heavily based on linear operators, we will adopt the notation where 2-tensors are written with the first index contravariant and the second index covariant, so they can be seen as linear operators. This allows us to drop the indices. With this convention, the first FMPD equation is written
\begin{equation}
\label{pdot_notations}
\frac{\widehat{\nabla} P}{d\tau} = \half R(S) \dot{X} + \frac{1}{2p} \cP(S) \frac{\widehat{\nabla} P}{d\tau}
\end{equation}

Now, this equation is of the form $\dot{P} = A + B \dot{P}$, which is solved as $\dot{P} = (\Id-B)^{-1} A$, provided the determinant of $\Id-B$ does not vanish. In other words, we have,
\begin{equation}
\frac{\widehat{\nabla} P}{d\tau} = \half \left(\Id_3 - \frac{1}{2p} \cP(S)\right)^{-1} R(S) \dot{X}
\end{equation}

We now need to calculate the inverse, and thus the determinant. For that, we can use the determinant formula $\epsilon_{\mu\nu\lambda} \det(M) = \epsilon_{\mu'\nu'\lambda'} M^{\mu'}{}_\mu M^{\nu'}{}_\nu M^{\lambda'}{}_\lambda$ to show that for square matrices in 3 dimensions,
\begin{equation}
\det(\Id_3 - M) = 1 - \Tr(M) - \det(M) + \half \left(\Tr(M)^2 - \Tr(M^2)\right)
\end{equation}
In our case, $M = \cP(S)/2p$, and the property that $\cP(S) l = 0$ implies $\det(M) = 0$, simplifying the calculations. This property also implies the following useful formula, obtained by computing $l^\lambda l_\sigma \epsilon^{\mu\nu\sigma} \epsilon_{\mu\nu\lambda} \det(M)$,
\begin{equation}
\Tr(\cP(S))^2 - \Tr(\cP(S)^2) = \frac{1}{s^2} S^{\mu\nu} S^{\mu'\nu'} \cP(S)^{\mu'}{}_\mu \cP(S)^{\nu'}{}_\nu
\end{equation}
Thus we have
\begin{subequations}
\label{result_det}
\begin{align}
\det\left(\Id_3 - \frac{1}{2p} \cP(S)\right) & = 1 - \frac{1}{2p} \cP(S)^\mu{}_\mu + \frac{1}{8p^2} \left(\left(\cP(S)^\mu{}_\mu\right)^2 - \cP(S)^\mu{}_\lambda \cP^\lambda{}_\mu\right)  \\
& = \frac{1}{p^2} \left(p^2 - \frac{p}{2} \cP(S)^\mu{}_\mu + \frac{1}{8s^2} \cP(S)_{\mu\nu} \cP(S)_{\lambda\rho} S^{\mu\lambda} S^{\nu\rho}\right) \\
& = \frac{1}{p^2} \widetilde{\Sigma}
\end{align}
\end{subequations}
Next, for the computation of the inverse, we need the adjugate matrix. There is again a useful explicit formula for 3-dimensional matrices,
\begin{equation}
\adj(M) = \half \left(\Tr(M)^2 - \Tr(M^2)\right) \Id_3 - M \Tr(M) + M^2 
\end{equation}
which we find can also be written in the form $\adj(M)^\rho{}_\sigma = \half \epsilon^{\rho\mu\nu} \epsilon_{\sigma\mu'\nu'} M^{\mu'}{}_\mu M^{\nu'}{}_\nu$. Finally, after calculating $\adj(\Id_3 - M) = (1 - \Tr(M)) \Id_3 + M + \adj(M)$, we find in our case where $\cP(S)l = 0$, which means that the only non-trivial component of $\adj(M)$ is its projection on $l_\rho l^\sigma$,
\begin{equation}
\label{result_adj}
\adj\left(\Id_3 - \frac{1}{2p} \cP(S)\right) =  \left(1 - \frac{\cP(S)^\lambda{}_\lambda}{2p} \right) \delta^\mu_\nu + \frac{1}{2p} \cP(S)^\mu{}_\nu + \frac{1}{8 p^2 s^2} \cP(S)_{\kappa\sigma} \cP(S)_{\lambda\rho} S^{\kappa\lambda} S^{\sigma\rho} l^\mu l_\nu
\end{equation}

On account of $\nabla P/d\tau = p \nabla l/d\tau$ and $l^2 = 1$, which imply $l^T R(S)\dot{X} = 0$ from \eqref{pdot_notations}, this gives us the explicit equation for $\dot{P}$,
\begin{align}
\frac{\widehat{\nabla} P}{d\tau} & = \frac{p}{2 \widetilde{\Sigma}} \left[\left(p - \half \Tr(\cP(S))\right) \Id_3 + \half \cP(S)\right] R(S)\dot{X}
\end{align}

Then, we can use the general product of 2 3-dimensional Levi-Civita tensors to calculate that, since $Sl = \cP(S)l = 0$,
\begin{equation}
S \cP(S) S = s^2 \cP(S)^T - s^2 \Tr(\cP(S))(\Id_3 - l l^T)
\end{equation}
which simplifies the momentum equation to,
\begin{equation}
\label{final_pdot1}
\frac{\widehat{\nabla} P}{d\tau} = \frac{p}{2 \widetilde{\Sigma}} \left[p \Id_3 + \frac{1}{2s^2} S \cP(S)^T S\right] R(S) \dot{X}
\end{equation}
We can use $S^2 = -s^2 (\Id_3 - l l^T)$ to bring the above equation to its final form
\begin{equation}
\label{final_pdot2}
\frac{\widehat{\nabla} P}{d\tau} = \frac{- p}{2 s^2 \widetilde{\Sigma}} S \left[p \Id_3 - \frac{1}{2} \cP(S)^T \right] S R(S) \dot{X}
\end{equation}

Now that we have an equation for the evolution of the momentum, it remains to obtain an equation for the velocity $\dot{X}$. To that end, we will work out a second way to obtain an explicit equation for the momentum and compare the two. Replacing $\frac{\widehat{\nabla} S}{d\tau}$ in the relation $\frac{\widehat{\nabla} S}{d\tau} P + S \frac{\widehat{\nabla} P}{d\tau} = 0$ by its expression \eqref{sdot_pydot}, we find,
\begin{equation}
\label{chd_qs_eq}
p\left(p (l^\mu l_\lambda - \delta^\mu_\lambda) + \half \cP(S)_\lambda{}^\mu \right) \dot{X}^\lambda + \left(S^\mu{}_\lambda + \half \widehat{Q}(S)^\mu{}_\lambda\right) \frac{\widehat{\nabla} P^\lambda}{d\tau} = 0,
\end{equation}
This equation is the same as Duval's equation \cite[(4.36)]{Duval08} up to a sign in front of $\widehat{Q}(S)$. The sign discrepancy comes from his unusual sign in the definition of the curvature, see above his \cite[(2.36)]{Duval08} which, in particular, changes the sign of $Q$ compared to \cite{BaoCS96}, the latter reference being the convention we use in this article.

Now, in dimension 3, skew symmetric operators are rather simple, in that their space is of dimension 3, hence we can characterize them by their kernel. In particular, we have the property that any skew symmetric 2-tensor $H$ whose kernel is generated by $l$ is proportional to $S$. This lets us calculate that, in 3 dimensions,
\begin{equation}
\widehat{Q}(S)^\mu{}_\nu = \frac{1}{2s^2} \widehat{Q}(S)(S) S^\mu{}_\nu
\end{equation}
where $\widehat{Q}(S)(S) = \widehat{Q}(S)_{\mu\nu} S^{\mu\nu}$. Putting the above expression in \eqref{chd_qs_eq} which we multiply on the left by $S$, and defining
\begin{equation}
\label{def_delta}
\Delta = s\left(1 + \frac{1}{4s^2} \widehat{Q}(S)(S)\right),
\end{equation}
we find,
\begin{equation}
\frac{\widehat{\nabla} P}{d\tau} = - \frac{p}{s \Delta} S\left[p \Id_3 - \half \cP(S)^T\right] \dot{X}
\end{equation}
Equating the above form for the derivative of $P$ with the previous form \eqref{final_pdot2}, and multiplying by $S$ on the left, we find,
\begin{equation}
\label{fancy_equation}
\left(\Id_3 - \frac{1}{2p} \cP(S)^T\right) \dot{X} - l (l \cdot \dot{X}) = \frac{\Delta}{2s \widetilde{\Sigma}} \left(\Id_3 - \frac{1}{2p} \cP(S)^T\right) S R(S) \dot{X} 
\end{equation}
We can multiply the above equation on the left by $(\Id_3 - \frac{1}{2p} \cP(S)^T)^{-1}$. Since $\cP(S) l = 0$, $l$ is clearly an eigenvector of $\left(\Id_3 - \frac{1}{2p} \cP(S)^T\right)$ with eigenvalue $1$, and hence it is also eigenvector of the inverse with eigenvalue $1$. This gives us the simple expression,
\begin{equation}
\label{equation_xdot}
\dot{X} - l (l \cdot \dot{X}) = \frac{\Delta}{2s \widetilde{\Sigma}} S R(S) \dot{X}
\end{equation}

Now, in dimension 3, $(l, S)$ form a complete basis, so we can decompose $\dot{X}$ in full generality as
\begin{equation}
\label{decomposition_xdot_3d}
\dot{X} = \alpha l + S V
\end{equation}
for some function $\alpha$ and some vector $V$ to be determined.

Putting \eqref{decomposition_xdot_3d} into \eqref{equation_xdot}, we find
\begin{equation}
SV = \alpha \frac{\Delta}{2s \widetilde{\Sigma}} SR(S)l + \frac{\Delta}{2s \widetilde{\Sigma}} S R(S) SV
\end{equation}

Now, to solve the above equation for $SV$, we need to simplify $S R(S) S V$. Since we are in the 3 dimensional case and since $S$ and $R(S)$ are skewsymmetric, they are each Hodge dual to a vector, and we can write $S_{\mu\nu} = \epsilon_{\mu\nu\lambda} s^\lambda$ and $R(S)_{\mu\nu} = \epsilon_{\mu\nu\lambda} J^\lambda$. A straightforward calculation then shows that $S R(S) S = \half \Tr(S R(S)) S = - \half R(S)(S) \, S$. Hence, we have,
\begin{equation}
2s \left(\frac{\widetilde{\Sigma}}{\Delta} + \frac{1}{4s} R(S)(S)\right) SV = \alpha SR(S)l
\end{equation}
By defining
\begin{equation}
\Sigma = \frac{\widetilde{\Sigma}}{\Delta} + \frac{1}{4s} R(S)(S)
\end{equation}
we have
\begin{equation}
\dot{X} = \alpha \left(l + \frac{1}{2s \Sigma} S R(S) l\right)
\end{equation}

Hence, in dimension 3, the complete equations of motion for the dynamical system described by the supplementary conditions $P = p l$ and $SP = 0$ are, upon choosing a suitable worldline parameter to set $\alpha = 1$,
\begin{subequations}
\label{eom_3d}
\begin{empheq}[box=\fbox]{align}
\dot{X} & = l + \frac{1}{2s \Sigma} S R(S) l \\
\frac{\widehat{\nabla} P}{d\tau} & = - \frac{p}{s \Delta} S\left[p \Id_3 - \half \cP(S)^T\right] \dot{X}
\end{empheq}
\end{subequations}

These are the equations of motion for a dynamical system with dipole moment $S$ in dimension 3. The notion of ``motion'' in this case is rather peculiar, in that there is no time involved, the motion is instantaneous. This is the case for instance for light in geometrical optics, or more relevant here, spinoptics akin to \cite{Duval08}, where a curve described by this kind of equations would correspond to a light ray in 3 dimensional space.

The above equations are the same equations of motion as obtained in \cite[(4.31)]{Duval08} using symplectic models, other than the sign in the definition of $\Delta$, which comes from the same remark below \eqref{chd_qs_eq}, giving us confidence in the methods developed in this article to obtain the FMPD equations and the supplementary conditions.

Note that in 3 dimensions, there is no equation for $S$ as this tensor is entirely determined by $P$ through \eqref{chd_def_p} and \eqref{chd_def_s}. The only degree of freedom is the value of the scalar spin, or absolute value of the dipole moment, $s$, and its sign.

\section{FMPD equations in Finsler spacetime}
\label{s:eom_4d}

The previous section dealt with dynamical systems in pure space, \textit{e.g.} determining the overall light ray in a particular crystal, since in geometric optics light travels with ``infinite velocity''. In most contexts however, it is more common to evaluate the motion of a dynamical system with respect to time, and thus consider spacetime, of dimension 3+1. In this section we will write the equations of motion in spacetime for both massive and massless dynamical systems.

\subsection{Massive dynamical systems}

The supplementary conditions we use here are essentially the same as in the previous section, 
\begin{subequations}
\begin{align}
P^\mu & = m l^\mu \label{mass_def_p} \\
S^\mu{}_\nu P^\nu & = 0 \label{mass_tul}
\end{align}
\end{subequations}
where the signature of the metric is such that $P^2 = m^2$, where $m$ is interpreted as the mass of the dynamical system, and we will still define the longitudinal spin $s$ (or more generally amplitude of the dipole moment) of the system as $\Tr(S^2) = - 2 s^2$.

Note that $S$ is a skewsymmetric operator, hence its rank is necessarily even. Since its kernel is at least of dimension 1 due to \eqref{mass_tul}, and if we assume $s \neq 0$, there exists another vector $J$ not parallel to $P$ such that $SJ = 0$. Note that $J$ is not unique, in that we can always shift it by $P$ without any effect on $S$. This vector is useful for our calculations later on, but our final result will not depend on $J$ (other than implicitly in the initial condition for $S$).
This vector can be chosen spacelike such that $P \cdot J = 0$ with the decomposition\footnote{This vector could be interpreted as encoding the vector components of the dipole moment. For example in Riemannian geometry, with the Minkowski metric, and $P = \partial_t$, one can set $(J^\mu) = (0, s_1, s_2, s_3)$ such that $S_{\mu\nu} = \begin{pmatrix}
0 & 0 & 0 & 0 \\
0 & 0 & -s_3 & s_2 \\
0 & s_3 & 0 & -s_1 \\
0 & -s_2 & s_1 & 0
\end{pmatrix}$ and where we would have $s_1^2 + s_2^2 + s_3^2 = s^2$.},
\begin{equation}
S_{\mu\nu} = \frac{s}{\sqrt{m^2 J^2}} \epsilon_{\mu\nu\lambda\rho} J^\lambda P^\rho
\end{equation}

With the above definitions, we can calculate a relation useful for later,
\begin{equation}
S^2 = -s^2 \Id + \frac{s^2}{m^2 J^2} \left( m^2 J J^T + J^2 P P^T \right)
\end{equation}

The vector $J$ means that now the decomposition of the velocity is, in full generality, 
\begin{equation}
\label{decomposition_xdot_4d}
\dot{X} = \alpha P + \beta J + SV
\end{equation}
for some $\alpha, \beta$ and $V$.

It is clear that the equations in 4 dimensions will be much more complicated than in 3 dimensions, the latter being already somewhat complicated. Moreover, it is most often the case that the dipole moment of the dynamical system is small compared to its other parameters. As a result, it makes sense to work out not the exact equations, but approximated equations, with the amplitude of the dipole moment $s$ small compared to other quantities of the same unit. With some abuse of physics notation, we will expand expressions in terms of $\cO(s^n)$.

With the above approximation in mind, the FMPD equations \eqref{FinslerMPD_cartan_py}, taking \eqref{mass_def_p} into account, are given by
\begin{subequations}
\label{FMPD_4d_expanded}
\begin{align}
\frac{\widehat{\nabla} P^\lambda}{d\tau} & = \half R(S)^\lambda{}_\nu \dot{X}^\nu + \frac{1}{2m}  \cP(S)^\lambda{}_\mu R(S)^\mu{}_\nu \dot{X}^\nu + \cO(s^3) \\
\frac{\widehat{\nabla} S^{\mu\lambda}}{d\tau} & = 2 P^{[\mu} \dot{X}^{\lambda]} + \dot{X}^\nu \cP(S)_\nu{}^{[\mu} l^{\lambda]} + \frac{1}{2m} l^{[\lambda} \widehat{Q}(S)^{\mu]}{}_\nu R(S)^\nu{}_\rho \dot{X}^\rho + \cO(s^3)
\end{align}
\end{subequations}
Then, injecting the above equations into $\frac{\widehat{\nabla} S}{d\tau} P + S \frac{\widehat{\nabla} P}{d\tau} = 0$ we get,
\begin{equation}
\begin{split}
0 & = P^\mu (\dot{X} \cdot P) - m^2 \dot{X}^\mu + \frac{m}{2} \dot{X}^\nu \cP(S)_\nu{}^\mu \\
& \quad + \frac{1}{4} \widehat{Q}(S)^\mu{}_\lambda R(S)^\lambda{}_\nu \dot{X}^\nu + \half S^\mu{}_\lambda R(S)^\lambda{}_\nu \dot{X}^\nu + \cO(s^3)
\end{split}
\end{equation}
Then, when introducing the decomposition of $\dot{X}$ \eqref{decomposition_xdot_4d},
\begin{equation}
\label{spdot_after_decomp}
\begin{split}
0 & = - m^2 (\beta J^\mu + S^\mu{}_\nu V^\nu) + \frac{m \beta}{2} J^\nu \cP(S)_\nu{}^\mu + \frac{m}{2} S^\nu{}_\lambda V^\lambda \cP(S)_\nu{}^\mu \\
& \quad + \frac{1}{4} \widehat{Q}(S)^\mu{}_\nu R(S)^\nu{}_\lambda \left(\alpha P^\lambda + \beta J^\lambda\right) + \frac{1}{2} S^\mu{}_\nu R(S)^\nu{}_\lambda \left(\alpha P^\lambda + \beta J^\lambda\right) + \cO(s^3)
\end{split}
\end{equation}
Contracting the above equation with $J_\mu$, we find,
\begin{equation}
\begin{split}
\cO(s^3) & = - m^2 \beta J^2 + \frac{m}{2} \beta J^\mu \cP(S)_{\mu\nu} J^\nu + \frac{m}{2} S^\nu{}_\lambda V^\lambda \cP(S)_\nu{}^\mu J_\mu  \\
& \quad + \frac{1}{4} J_\mu \widehat{Q}(S)^\mu{}_\nu R(S)^\nu{}_\lambda \left(\alpha P^\lambda + \beta J^\lambda\right)
\end{split}
\end{equation}
which is solved for $\beta$ as
\begin{equation}
\beta = \frac{1}{2m^2 J^2} \left(\frac{\alpha}{2} J_\mu \widehat{Q}(S)^\mu{}_\nu R(S)^\nu{}_\lambda P^\lambda + m S^\nu{}_\lambda V^\lambda \cP(S)_\nu{}^\mu J_\mu \right) + \cO(s^3)
\end{equation}
In particular, we see that $\beta$ is of order $s^2$, meaning several terms drop out of \eqref{spdot_after_decomp},
\begin{equation}
\begin{split}
\label{spdot_after_beta}
\cO(s^3) & = - m^2 (\beta J^\mu + S^\mu{}_\nu V^\nu) + \frac{m}{2} S^\nu{}_\lambda V^\lambda \cP(S)_\nu{}^\mu \\
& \quad + \frac{\alpha}{4} \widehat{Q}(S)^\mu{}_\nu R(S)^\nu{}_\lambda P^\lambda  + \frac{\alpha}{2} S^\mu{}_\nu R(S)^\nu{}_\lambda P^\lambda
\end{split}
\end{equation}
From the above equation, we see that the only term seemingly of order $s$ is $m^2 S^\mu{}_\nu V^\nu$. This means that $V$ is necessarily of order at least $s$, thus simplifying $\beta$,
\begin{equation}
\beta = \frac{\alpha}{4m^2 J^2} J_\mu \widehat{Q}(S)^\mu{}_\nu R(S)^\nu{}_\lambda P^\lambda + \cO(s^3) 
\end{equation}
as well as the equation \eqref{spdot_after_beta} which can then immediately be solved as
\begin{equation}
\beta J^\mu + S^\mu{}_\nu V^\nu = \frac{\alpha}{4m^2} \widehat{Q}(S)^\mu{}_\nu R(S)^\nu{}_\lambda P^\lambda + \frac{\alpha}{2m^2} S^\mu{}_\nu R(S)^\nu{}_\lambda P^\lambda + \cO(s^3) 
\end{equation}
which we can insert into the decomposition of the velocity \eqref{decomposition_xdot_4d},
\begin{equation}
\dot{X} = \alpha P^\mu + \frac{\alpha}{2m^2} \left(S^\mu{}_\nu R(S)^\nu{}_\lambda P^\lambda + \half \widehat{Q}(S)^\mu{}_\nu R(S)^\nu{}_\lambda P^\lambda \right) + \cO(s^3)
\end{equation}

Finally, we can write the velocity relation in the FMPD equations \eqref{FMPD_4d_expanded}, and after choosing a worldline parameter such that $\alpha = 1$ we finally obtain,
\begin{subequations}
\label{eom_4d_massive}
\begin{empheq}[box=\fbox]{align}
\dot{X} & = P^\mu + \frac{1}{2m^2} \left(S^\mu{}_\nu R(S)^\nu{}_\lambda P^\lambda + \half \widehat{Q}(S)^\mu{}_\nu R(S)^\nu{}_\lambda P^\lambda \right) + \cO(s^3) \\
\frac{\widehat{\nabla} P^\lambda}{d\tau} & = \half R(S)^\lambda{}_\nu P^\nu + \frac{1}{4m}  \cP(S)^\lambda{}_\mu R(S)^\mu{}_\nu P^\nu + \cO(s^3) \\
\frac{\widehat{\nabla} S^{\mu\lambda}}{d\tau} & = \frac{1}{m^2} P^{[\mu} S^{\lambda]}{}_\rho R(S)^\rho{}_\sigma P^\sigma + \cO(s^3)
\end{empheq}
\end{subequations}

Conveniently, these equations do not contain the vector $J$ explicitly. 

\subsection{Massless dynamical systems}

We finally turn our attention to massless dynamical systems. It turns out that they are conceptually harder to grasp, but technically easier to handle than the previous massive systems. This is because the apparent motion of massless particles with spin is observer dependent, due to Wigner-Souriau translations \cite{Souriau70,StoneDZ14}. While the same is true for massive particles, the difference between trajectories is bounded and not meaningful \cite{CostaN14}. But for massless particles, the difference between observed trajectories is unbounded \cite{HarteO22}. 

Obviously, the main difference between massless and massive dynamical system is the momentum, in that we need $P^2 = 0$, which means that $P$ cannot be proportional to $l$. Instead, we can set the momentum as the Finsler variable $y$ directly, without normalization by $F$. Hence, our supplementary conditions here are
\begin{subequations}
\begin{align}
P^\mu & = y^\mu \label{massless_def_p} \\
S^\mu{}_\nu P^\nu & = 0 \label{massless_tul}
\end{align}
\end{subequations}
Note however that we are now working on a slightly different space than the one used to obtain the equations of motion \eqref{FinslerMPD_chern}, since we cannot use the distinguished section $l$ or other objects rescaled by $F$ since here $F^2 = P^2 = 0$ on shell. This means that for photons, we need to use the inhomogeneous FMPD equations \eqref{FinslerMPD_cartan_inhomogeneous}.

Like in the massive case, the dipole moment tensor $S$ is skewsymmetric and hence there is a second vector, independent from $P$ generating its 2-dimensional kernel. Now, this second vector is somewhat controversial in the massless case, in that there are (at least) two seemingly different schools of thoughts about it.

One approach, similar to that of the previous section, is to derive the equations of motion by working solely from the relation $SP = 0$. While in the massless case, the interpretation of the second vector, which we will call $t$\footnote{Note that $t$ is not unique, since we are free to shift it by $P$.}, is not straightforward, the equations we obtain in this way do not have an explicit dependency on $t$. There is however a dependency in the initial condition for $S$, since it contains $t$, which becomes a problem as to how $t$ and thus $S$ should be interpreted and defined initially.

The second approach, much more recent, see \cite{OanceaJDRPA20,HarteO22}, comes from WKB expansion of Maxwell equations in curved spacetime, where the authors realized that a second vector generating the kernel of $S$ is the velocity $t$ of a local timelike observer. This is linked to the fact that due to so-called Souriau-Wigner translations, the trajectory of massless dynamical systems with spin is indeed observer dependent \cite{StoneDZ14}. They then derive equations of motion from the relation $St = 0$, keeping the dependency on the observer explicit. This approach has the obvious advantage of understanding where the second vector comes from, which makes interpretation and definition of initial conditions much more straightforward.

In both cases, since $P$ and $t$ generate the kernel of $S$, and $P^2 = 0$, we have the decomposition,
\begin{equation}
\label{massless_def_s}
S_{\mu\nu} = \frac{s}{P \cdot t} \epsilon_{\mu\nu\lambda\sigma} t^\lambda P^\sigma
\end{equation}
such that $P \cdot t \neq 0$.

Now let us consider the first approach described above. We differentiate the usual Tulczyjew condition $SP = 0$, and we get a rather simple expression on account of $P^2 = 0$,
\begin{equation}
\left(\dot{X} \cdot P\right) P + \half S R(S) \dot{X} + \frac{1}{2p} S\cP_I(S) \frac{\widehat{\nabla} P}{d\tau} = 0
\end{equation}
Contracting the above equation with $t_\mu$, we find $(\dot{X} \cdot P) (P \cdot t) = 0$. Since $P \cdot t \neq 0$, we have $\dot{X} \cdot P = 0$. In general, we can again decompose the velocity in full generality as $\dot{X}^\mu = \alpha P^\mu + \beta t^\mu + S^\mu{}_\nu V^\nu$ for some $\alpha$, $\beta$ and $V$. However $\dot{X} \cdot P = 0$ immediately implies $\beta = 0$. We then choose the worldline parameter such that $\alpha = 1$, or equivalently $\dot{X} \cdot t = P \cdot t$, so that,
\begin{equation}
\label{massless_decomp_xdot}
\dot{X}^\mu = P^\mu + S^\mu{}_\nu V^\nu
\end{equation}

Using this decomposition in the previous equation, we get
\begin{equation}
S R(S) P + S R(S) SV + \frac{1}{p} S\cP_I(S) \frac{\widehat{\nabla} P}{d\tau} = 0
\end{equation}

We again need to simplify $S R(S) S V$. This is however slightly more complicated in 4 dimensions. In general, for any skew-symmetric operators $\Omega$ and $F$,
\begin{equation}
\label{magic}
\Omega F \Omega = \Pf(\Omega) \star(F) + \half \Tr(\Omega F) \Omega
\end{equation}
where $\Pf$ is the pfaffien, \textit{i.e.} the square root of the determinant, and $\star$ the Hodge star.

Although this formula is very useful when working with the MPD equations, its proof is in \cite{Duval72} which is unpublished and unavailable online. It can be proven as follows. In 4 dimensions, the pfaffian's expression is $\Pf(\Omega) = \frac{1}{8} \epsilon^{\mu\nu\lambda\sigma} \Omega_{\mu\nu} \Omega_{\lambda\sigma}$. We can multiply both sides by $\epsilon_{\mu'\nu'\lambda'\sigma'}$ and expand the Levi Civata tensor product to find the totally skewsymmetric $\epsilon_{\mu\nu\lambda\sigma} \Pf(\Omega) = \Omega_{[\mu\nu} \Omega_{\lambda\sigma]} = \Omega_{\mu\nu} \Omega_{\lambda \sigma} + \Omega_{\mu\lambda} \Omega_{\sigma\nu} + \Omega_{\mu \sigma} \Omega_{\nu\lambda}$. Contraction with $\half F^{\lambda\sigma}$ then yields \eqref{magic}.

Due to \eqref{magic}, we find just as in the 3-dimensional case $SR(S)S = - \half R(S)(S) \, S$, and,
\begin{equation}
S R(S) P -\half R(S)(S) SV + \frac{1}{p} S\cP_I(S) \frac{\widehat{\nabla} P}{d\tau} = 0
\end{equation}
which is immediate to solve for $SV$ assuming $R(S)(S) \neq 0$, and implies, for the velocity equation,
\begin{equation}
\dot{X}^\mu = P^\mu + \frac{2}{R(S)(S)} \left(S^\mu{}_\nu R(S)^\nu{}_\lambda P^\lambda + \frac{1}{p} S^\mu{}_\nu \cP_I(S)^\nu{}_\lambda \frac{\widehat{\nabla} P^\lambda}{d\tau} \right)
\end{equation}

Inserting this velocity relation into the FMPD equation for $\frac{\widehat{\nabla} P^\lambda}{d\tau}$, and using the formula \eqref{magic} to simplify $R(S) S R(S)$, we find,
\begin{equation}
\frac{\widehat{\nabla} P}{d\tau} = s \frac{\Pf(R(S))}{R(S)(S)} P + \frac{1}{2p} \left(\Id + \frac{2}{R(S)(S)} R(S) S\right) \cP_I(S) \frac{\widehat{\nabla} P}{d\tau}
\end{equation}

Since $\frac{\widehat{\nabla} S}{d\tau} P = 0$, we have $S \frac{\widehat{\nabla} P}{d\tau} = 0$ \textit{i.e.} $\frac{\widehat{\nabla} P}{d\tau} = \alpha P + \beta t$ for some $\alpha$ and $\beta$. Then, $P^2 = 0$ and $P\cdot t \neq 0$ imply $\beta = 0$ and we find the derivative of the momentum is simply proportional to the momentum, $\widehat{\nabla} P^\lambda/d\tau = \alpha P^\lambda$. Injecting this in the above equation, we immediately have, since $\cP(S) P = 0$,
\begin{equation}
\frac{\widehat{\nabla} P^\lambda}{d\tau} = s \frac{\Pf(R(S))}{R(S)(S)} P^\lambda
\end{equation}

This can be injected into the velocity relation, and the final set of equations for massless particles with dipole moment takes a simple form,
\begin{subequations}
\label{eom_4d_massless_exact}
\begin{empheq}[box=\fbox]{align}
\dot{X}^\mu & = P^\mu + \frac{2}{R(S)(S)} S^\mu{}_\nu R(S)^\nu{}_\lambda P^\lambda \\
\frac{\widehat{\nabla} P^\lambda}{d\tau} & = s \frac{\Pf(R(S))}{R(S)(S)} P^\lambda \label{eom_4d_massless_exact_dp} \\
\frac{\widehat{\nabla} S^{\mu\lambda}}{d\tau} & = 2 P^{[\mu} \dot{X}^{\lambda]} + \dot{X}^\nu \cP_I(S)_\nu{}^{[\mu} P^{\lambda]} \label{eom_4d_massless_exact_ds}
\end{empheq}
\end{subequations}

Remarkably, the first two equations for the position and momentum have no pure Finsler contribution other than in the definitions of the derivatives and the Riemann tensor. This makes the equations look very similar to the MPD equations applied to photons, sometimes called the Souriau-Saturnini equations \cite{Saturnini76,DuvalS16}, to which they reduce to when the geometry is Riemannian.

These equations share the same apparent problems as the Souriau-Saturnini equations. The first one being that the second term in the velocity equation for $\dot{X}$ is the ratio of two small numbers, which makes the equations particularly difficult to solve both analytically, and numerically \cite{DuvalMS18}.

Another obvious problem concerns cases when the Riemann tensor vanishes, for example for flat spacetime. The equations \eqref{eom_4d_massless_exact} were derived on the assumption that $R(S)(S) \neq 0$, so they are not valid in such case. If the Riemann tensor vanishes, it is impossible to obtain equations of motion that do not depend on the velocity $t$ of the observer.

Overall, while the equations \eqref{eom_4d_massless_exact} are exact and do not involve the observer explicitly, we will see that the equations of motion obtained from the second method are more convenient to use in practice, since they are free of these two issues.

\bigskip

We now move on to the second approach mentioned at the beginning of the subsection. Since we want the vector $t$ explicitly in the equations, we will differentiate the Corinaldesi-Papapetrou condition \cite{CorinaldesiP51} $S^\mu{}_\nu t^\nu = 0$ instead of the usual Tulczyjew condition. We emphasis that using the Corinaldesi-Papapetrou condition does not mean that the Tulczyjew condition does not hold, both hold at the same time due to the even dimension of spacetime and the definition of $S$ \eqref{massless_def_s}. 

From $\frac{\widehat{\nabla} S^\mu{}_\nu}{d\tau} t^\nu + S^\mu{}_\nu \frac{\widehat{\nabla} t^\nu}{d\tau} = 0$, we readily obtain,
\begin{equation}
\label{dst_sdt}
\begin{split}
0 & =  P^\mu (\dot{X} \cdot t) - \dot{X}^\mu (P\cdot t)  + \frac{P \cdot t}{2} \dot{X}^\nu \cP_I(S)_\nu{}^\mu - \frac{1}{2} \dot{X}^\nu \cP_I(S)_\nu{}^\lambda t_\lambda P^\mu  \\
& \quad + \frac{P\cdot t}{2} \widehat{Q}_I(S)^\mu{}_\nu \frac{\widehat{\nabla} P^\nu}{d\tau} - \frac{1}{2} t_\lambda \widehat{Q}_I(S)^\lambda{}_\nu \frac{\widehat{\nabla} P^\nu}{d\tau} P^\mu + S^\mu{}_\nu \frac{\widehat{\nabla} t^\nu}{d\tau}	
\end{split}
\end{equation}

We have seen previously that $\frac{\widehat{\nabla} P^\nu}{d\tau}$ is proportional to $P$, which means that the terms of the form $\widehat{Q}_I(S)\frac{\widehat{\nabla} P}{d\tau}$ vanish. Then, putting the decomposition of $\dot{X}$ \eqref{massless_decomp_xdot} in the above equation, we get,
\begin{align}
S^\mu{}_\nu V^\nu & = \frac{1}{P\cdot t} S^\mu{}_\nu \frac{\widehat{\nabla} t^\nu}{d\tau}	 + \frac{1}{2} S^\nu{}_\rho V^\rho \cP_I(S)_\nu{}^\lambda \left(\delta^\mu_\lambda - \frac{P^\mu t_\lambda}{P \cdot t}\right) \\
& =  \frac{-1}{2(P \cdot t)^2} S^\nu{}_\rho \frac{\widehat{\nabla} t^\rho}{d\tau} \cP_I(S)_\nu{}^\lambda t_\lambda P^\mu + \frac{1}{P\cdot t} \left(\delta^\mu_\lambda - \half \cP_I(S)_\lambda{}^\mu\right)^{-1} S^\lambda{}_\rho \frac{\widehat{\nabla} t^\rho}{d\tau}
\end{align}
provided $\det(\delta^\mu_\lambda - \half \cP_I(S)_\lambda{}^\mu) \neq 0$. This means that the velocity is given by,
\begin{equation}
\label{eom_Xdot_massless_observer}
\begin{split}
\dot{X}^\mu & = P^\mu - \frac{1}{2(P \cdot t)^2} S^\nu{}_\rho \frac{\widehat{\nabla} t^\rho}{d\tau} \cP_I(S)_\nu{}^\lambda t_\lambda P^\mu + \frac{1}{P\cdot t} \left(\delta^\mu_\lambda - \half \cP_I(S)_\lambda{}^\mu\right)^{-1} S^\lambda{}_\rho \frac{\widehat{\nabla} t^\rho}{d\tau}
\end{split}
\end{equation}

The final equations of motion are then,
\begin{subequations}
\label{eom_4d_massless_obs}
\begin{empheq}[box=\fbox]{align}
\dot{X}^\mu & = P^\mu - \frac{S^\nu{}_\rho}{2(P \cdot t)^2}  \frac{\widehat{\nabla} t^\rho}{d\tau} \cP_I(S)_\nu{}^\lambda t_\lambda P^\mu + \frac{1}{P\cdot t} \left(\delta^\mu_\lambda - \half \cP_I(S)_\lambda{}^\mu\right)^{-1} S^\lambda{}_\rho \frac{\widehat{\nabla} t^\rho}{d\tau} \label{eom_4d_massless_obs_dx} \\
\frac{\widehat{\nabla} P^\mu}{d\tau} & = \half \left(\delta^\mu_\lambda - \half \cP_I(S)^\mu{}_\lambda\right)^{-1} R(S)^\lambda{}_\nu \dot{X}^\nu \label{eom_4d_massless_obs_dp} \\
\frac{\widehat{\nabla} S^{\mu\lambda}}{d\tau} & = 2 P^{[\mu} \dot{X}^{\lambda]} + \dot{X}^\nu \cP_I(S)_\nu{}^{[\mu} P^{\lambda]}  \label{eom_4d_massless_obs_ds}
\end{empheq}
\end{subequations}

While the inverses could be calculated explicitly, the formulas get rather complicated in 4 dimensions\footnote{For a singular, 4 dimensional, square matrix $M$, we calculate
\begin{equation*}
(\Id-M)^{-1} = \Id + \frac{\left(1 - \Tr(M) + \half \left(\Tr(M)^2 - \Tr(M^2)\right)\right) M + (1 - \Tr(M)) M^2 + M^3}{1 - \Tr M + \half \left(\Tr(M)^2 - \Tr(M^2)\right) - \frac{1}{6}\left(\Tr(M)^3 - 3 \Tr(M) \Tr(M^2) + 2 \Tr(M^3)\right)}
\end{equation*}} and it seems more convenient to calculate it once a specific metric has been chosen.

The equations \eqref{eom_4d_massless_obs} seem to have different properties than those in \eqref{eom_4d_massless_exact} at first sight, in particular regarding rescalings of $s$. The velocity equation in \eqref{eom_4d_massless_exact} seems invariant under rescalings of $s$. While the velocity equation of \eqref{eom_4d_massless_obs} does not seem to be, both equations share the same rescaling property. Actually, rescaling $s$ while keeping other parameters fixed, in particular lengths, mean rescaling $\hbar$. Since $P \cdot t$ is (minus) the energy of the particle, rescaling $\hbar$ while keeping lengths fixed rescales this quantity, and in general rescales $P$. The same is true for $\cP(S)$, since $\cP$ is a derivative of the metric by what is in the end the momentum, $\cP(S)$ turns out to be invariant under rescalings of $\hbar$. Moreover, $\frac{\widehat{\nabla} t^\nu}{d\tau} = \dot{X}^\lambda t^\nu{}_{|\lambda} + \dot{Y}^\lambda t^\nu{}_{;\lambda}$ rescales like the momentum $P$. In the end, we find that \eqref{eom_4d_massless_obs_dx} rescales like $P$, just like \eqref{eom_4d_massless_exact}.

As mentioned previously, the equations of motion \eqref{eom_4d_massless_obs} conveniently do not degenerate in the flat case or when $R(S)(S) = 0$ in general. This means that using these equations is a must if one calculates a trajectory where $R(S)(S)$ vanishes.

However, there are still some points that need to be considered carefully with these equations. The first point is that when deriving these equations exactly from the FMPD equations, one is not free to choose the observer velocity $t$. The third equation on the evolution of $S$ \eqref{eom_4d_massless_obs_ds} and the decomposition of $S$ \eqref{massless_def_s} offer a constraint on $t$. In particular, such compatibility relations show that $t$ must be chosen so that $\frac{\widehat{\nabla} P^\mu}{d\tau} \propto P$, matching \eqref{eom_4d_massless_exact_dp}. This is unlike in a WKB expansion such as in \cite{HarteO22} (for the MPD equations) where, with the expansion, only the first two equations \eqref{eom_4d_massless_obs_dx}, \eqref{eom_4d_massless_obs_dp} appear and the observer is freely chosen. 

A second point to look out for is that $\frac{\widehat{\nabla} t^\nu}{d\tau} = \dot{X}^\lambda t^\nu{}_{|\lambda} + \dot{Y}^\lambda t^\nu{}_{;\lambda}$, which means that some operator needs to be inverted in order to finally compute $\dot{X}$, and this operation may lead to some complicated expression.

\section{Discussions}

In this article, we have applied Souriau's Principle of General Covariance, which allows to find diffeomorphism invariant equations of motion for particles with multipole moments, to Finsler geometry. The equations of motion we found \eqref{FinslerMPD_cartan} are the generalization of the Mathisson-Papapetrou-Dixon equations to Finsler geometries. While the equations involve Finsler objects, the conserved quantities we derived \eqref{cons_quant} for these equations are formally the same as in the Riemannian case, though not exactly the same since a covariant derivative appears in the expression, which is slightly different in the Finsler case.

While the derivation of the FMPD equations \eqref{FinslerMPD_cartan} is straightforward, some interpretations are needed for physical applications. The first point is what vector of the dynamical system should the $y$ coordinate of Finsler manifolds be? While the question needs not be asked when considering Finslerian geodesics, since in that case the velocity $\dot{X}$ and the momentum $P$ are parallel, the FMPD equations predict that these two vectors are different in general. In this article, we argued that the momentum is more natural, since it is a vector that defines a state on phase space, unlike the velocity, and since the equations of motion take a more ``natural'' and simpler look.

In a similar fashion to the standard MPD equations, one needs supplementary conditions to close the system of equations. By using the Tulczyjew condition, we wrote the closed equations of motion for a dynamical system in 3 spatial dimensions, which is the space to consider to calculate the light rays in geometrical optics, which do propagate instantaneously. In doing so, the equations of motion \eqref{eom_3d} we obtained are the same as in \cite{Duval08} for Finsler spinoptics, where they were obtained in an independent way using symplectic models. This comparison gives confidence in the techniques and supplementary conditions we adopted in the present paper.

Then, using the same supplementary condition as in the previous case, we tackled writing equations of motion on a 4 dimensional spacetime. There are different classes of dynamical systems in 4 dimensions, namely massive and massless ones (and tachyonic ones which we did not consider here). As 4 dimensional equations are substantially more complicated, we only offer an expansion of the equations in the massive case when the spin is small (compared to other scales of the system) \eqref{eom_4d_massive}. Massless equations of motion are more subtle as it is known the trajectories are observer dependent in an unbounded way \cite{StoneDZ14}. We give two ways of writing the equations of motion. One is on the model of the Riemannian construction in \cite{OanceaJDRPA20} where the equations are explicitly dependent on the evolution of the observer on the worldline \eqref{eom_4d_massless_obs}, and the other one where said evolution is ``backed in'' the equations \eqref{eom_4d_massless_exact}, giving rise to equations which do not contain the observer explicitly. As in the Riemannian case, we argued that the explicitly observer dependent equations of motion should be easier to work with in practice.

These equations, in 3 dimensions for optics, and 4 dimensions for massive or massless systems, can be used to study more precisely the motion of dynamical systems with angular momentum, in Finsler spacetimes. This can be useful in two ways. First, there are currently some works studying motion of photons in Finsler spacetimes, in a way that is trying to find small deviations to geodesics and deducing constraints to the Finsler parameters of the spacetime \cite{Shen23}. However photons do have spin, which does also introduce deviations to geodesics \cite{DuvalMS18,OanceaJDRPA20}. Using the FMPD equations which do take both effects into considerations, allows to attribute the experimental constraints correctly. Second, there might be some interesting physics happening in the coupling of dipole moments and Finsler parameters, which could be worth studying in their own right.

\section*{Acknowledgments}

We would like to thank Tiberu Harko for discussions, and Yuhong Fang for great advice in the early stage of this project. YL would also like to thank Yongsheng Huang for his encouragement to pursue in this field and for his general support during this research.

Funding: This work is supported by the National Natural Science Foundation of China (Grant No. W2433003).


\begin{thebibliography}{10}

\bibitem{Frenkel26}
J.~Frenkel, ``Die Elektrodynamik des rotierenden Elektrons'',
  \href{https://dx.doi.org/https://doi.org/10.1007/BF01397099 }{ Z. Phys. {\bf
  37}, p.~243, (1926)}.

\bibitem{Mathisson37}
M.~Mathisson, ``Neue Mechanik materieller Systeme'', Acta Phys. Pol. {\bf 6},
  p.~163 (1937),
  \url{http://delibra.bg.polsl.pl/dlibra/publication/39562/edition/34807/content}.

\bibitem{Papapetrou51}
A.~Papapetrou, ``Spinning Test-Particles in General Relativity. I'',
  \href{https://dx.doi.org/https://doi.org/10.1098/rspa.1951.0200 }{ Proc. R.
  Soc. A {\bf 209}, p.~248, (1951)}.

\bibitem{Pirani56}
F.~A.~E. Pirani, ``On the physical significance of the Riemann tensor'', Acta
  Phys. Pol. {\bf 15}, p.~389 (1956).
\newblock Republication: On the physical significance of the Riemann tensor,
  \href{https://doi.org/10.1007/s10714-009-0787-9}{Gen. Relativ. Gravit. {\bf
  41}, p.~1215, (2019)}.

\bibitem{Tulczyjew59}
W.~Tulczyjew, ``Motion of multipole particles in general relativity theory'',
  Acta Phys. Pol. {\bf 18}, p.~393 (1959).

\bibitem{Dixon70}
W.~G. Dixon, ``Dynamics of Extended Bodies in General Relativity. I. Momentum
  and Angular Momentum'',
  \href{https://dx.doi.org/https://doi.org/10.1098/rspa.1970.0020 }{ Proc. R.
  Soc. A {\bf 314}, p.~499, (1970)}.

\bibitem{CostaN14}
L.~F.~O. Costa, J.~Nat\'ario, ``{Center of mass, spin supplementary conditions,
  and the momentum of spinning particles}'',
  \href{https://dx.doi.org/10.1007/978-3-319-18335-0_6 }{ Fund. Theor. Phys.
  {\bf 179}, p.~215, (2015)}, arXiv:
  \href{https://arxiv.org/abs/1410.6443}{1410.6443}.

\bibitem{PlyatskoSF11}
R.~Plyatsko, O.~Stefanyshyn, M.~Fenyk, ``{Mathisson-Papapetrou-Dixon equations
  in the Schwarzschild and Kerr backgrounds}'',
  \href{https://dx.doi.org/10.1088/0264-9381/28/19/195025 }{ Class. Quant.
  Grav. {\bf 28}, p.~195025, (2011)}, arXiv:
  \href{https://arxiv.org/abs/1110.1967}{1110.1967}.

\bibitem{BiniGV17}
D.~Bini, A.~Geralico, J.~Vines, ``{Hyperbolic scattering of spinning particles
  by a Kerr black hole}'', \href{https://dx.doi.org/10.1103/PhysRevD.96.084044
  }{ Phys. Rev. D {\bf 96}, p.~084044, (2017)}, arXiv:
  \href{https://arxiv.org/abs/1707.09814}{1707.09814}.

\bibitem{SkoupyL21}
V.~Skoup\'y, G.~Lukes-Gerakopoulos, ``{Spinning test body orbiting around a
  Kerr black hole: Eccentric equatorial orbits and their asymptotic
  gravitational-wave fluxes}'',
  \href{https://dx.doi.org/10.1103/PhysRevD.103.104045 }{ Phys. Rev. D {\bf
  103}, p.~104045, (2021)}, arXiv:
  \href{https://arxiv.org/abs/2102.04819}{2102.04819}.

\bibitem{LadinoBLRA23}
J.~M. Ladino, C.~A. Benavides-Gallego, E.~Larra\~naga, J.~Rayimbaev,
  F.~Abdulxamidov, ``{Charged spinning and magnetized test particles orbiting
  quantum improved charged black holes}'',
  \href{https://dx.doi.org/10.1140/epjc/s10052-023-12187-2 }{ Eur. Phys. J. C
  {\bf 83}, p.~989, (2023)}, arXiv:
  \href{https://arxiv.org/abs/2305.15350}{2305.15350}.

\bibitem{DuvalS16}
C.~Duval, T.~Sch{\"u}cker, ``{Gravitational birefringence of light in
  Robertson-Walker cosmologies}'',
  \href{https://dx.doi.org/10.1103/PhysRevD.96.043517 }{ Phys. Rev. D {\bf 96},
  p.~043517, (2017)}, arXiv:
  \href{https://arxiv.org/abs/1610.00555}{1610.00555}.

\bibitem{DuvalMS18}
C.~Duval, L.~Marsot, T.~Sch{\"u}cker, ``{Gravitational birefringence of light
  in Schwarzschild spacetime}'',
  \href{https://dx.doi.org/10.1103/PhysRevD.99.124037 }{ Phys. Rev. D {\bf 99},
  p.~124037, (2019)}, arXiv:
  \href{https://arxiv.org/abs/1812.03014}{1812.03014}.

\bibitem{Vines18}
J.~Vines, ``{Scattering of two spinning black holes in post-Minkowskian
  gravity, to all orders in spin, and effective-one-body mappings}'',
  \href{https://dx.doi.org/10.1088/1361-6382/aaa3a8 }{ Class. Quant. Grav. {\bf
  35}, p.~084002, (2018)}, arXiv:
  \href{https://arxiv.org/abs/1709.06016}{1709.06016}.

\bibitem{ChenCHK22}
W.-M. Chen, M.-Z. Chung, Y.-t. Huang, J.-W. Kim, ``{The 2PM Hamiltonian for
  binary Kerr to quartic in spin}'',
  \href{https://dx.doi.org/10.1007/JHEP08(2022)148 }{ JHEP {\bf 08}, p.~148,
  (2022)}, arXiv: \href{https://arxiv.org/abs/2111.13639}{2111.13639}.

\bibitem{GosselinBM06}
P.~Gosselin, A.~Berard, H.~Mohrbach, ``{Spin Hall effect of photons in a static
  gravitational field}'', \href{https://dx.doi.org/10.1103/PhysRevD.75.084035
  }{ Phys. Rev. D {\bf 75}, p.~084035, (2007)}, arXiv:
  \href{https://arxiv.org/abs/hep-th/0603227}{hep-th/0603227}.

\bibitem{OanceaJDRPA20}
M.~A. Oancea, J.~Joudioux, I.~Dodin, D.~Ruiz, C.~F. Paganini, L.~Andersson,
  ``{Gravitational spin Hall effect of light}'',
  \href{https://dx.doi.org/10.1103/PhysRevD.102.024075 }{ Phys. Rev. D {\bf
  102}, p.~024075, (2020)}, arXiv:
  \href{https://arxiv.org/abs/2003.04553}{2003.04553}.

\bibitem{Har74}
M.~Harwit, R.~V.~E. Lovelace, B.~Dennison, D.~L. Jauncey, J.~Broderick,
  ``Gravitational deflection of polarised radiation'',
  \href{https://dx.doi.org/https://doi.org/10.1038/249230a0 }{ Nature {\bf
  249}, p.~230, (1974)}.

\bibitem{LianZ24}
D.-D. Lian, P.-M. Zhang, ``{The motion of twisted particles in a stellar
  gravitational field}'', \href{https://dx.doi.org/10.1088/1361-6382/ad721d }{
  Class. Quant. Grav. {\bf 41}, p.~195007, (2024)}, arXiv:
  \href{https://arxiv.org/abs/2312.14391}{2312.14391}.

\bibitem{Imbert72}
C.~Imbert, ``Calculation and Exprimental Proof of the Transverse Shift Induced
  by Total Internal Reflection of a Circularly Polarized Light Beam'',
  \href{https://dx.doi.org/10.1103/PhysRevD.5.787 }{ Phys. Rev. D {\bf 5},
  p.~787, (1972)}.

\bibitem{BliokhNKH08}
K.~Y. Bliokh, A.~Niv, V.~Kleiner, E.~Hasman, ``{Geometrodynamics of Spinning
  Light}'', \href{https://dx.doi.org/10.1038/nphoton.2008.229 }{ Nat. Photonics
  {\bf 2}, p.~748, (2008)}, arXiv:
  \href{https://arxiv.org/abs/0810.2136}{0810.2136}.

\bibitem{HostenK08}
O.~Hosten, P.~Kwiat, ``Observation of the Spin Hall Effect of Light via Weak
  Measurements'', \href{https://dx.doi.org/10.1126/science.1152697 }{ Science
  {\bf 319}, p.~787, (2008)}.

\bibitem{Clayton15}
J.~D. Clayton, ``{On Finsler Geometry and Applications in Mechanics: Review and
  New Perspectives}'', \href{https://dx.doi.org/10.1155/2015/828475 }{ Advances
  in Mathematical Physics {\bf 2015}, p.~828475, (2015)}.

\bibitem{AntonelliIM93}
P.~L. Antonelli, R.~S. Ingarden, M.~Matsumoto, {\em The Theory of Sprays and
  Finsler Spaces with Applications in Physics and Biology}.
\newblock Kluwer Academic Press, Dordrecht, 1993, doi:
  \href{https://doi.org/10.1007/978-94-015-8194-3}{10.1007/978-94-015-8194-3}.

\bibitem{ChangL08}
Z.~Chang, X.~Li, ``{Modified Newton's gravity in Finsler Space as a possible
  alternative to dark matter hypothesis}'',
  \href{https://dx.doi.org/10.1016/j.physletb.2008.09.010 }{ Phys. Lett. B {\bf
  668}, p.~453, (2008)}, arXiv:
  \href{https://arxiv.org/abs/0806.2184}{0806.2184}.

\bibitem{HamaHSS21}
R.~Hama, T.~Harko, S.~V. Sabau, S.~Shahidi, ``{Cosmological evolution and dark
  energy in osculating Barthel\textendash{}Randers geometry}'',
  \href{https://dx.doi.org/10.1140/epjc/s10052-021-09517-7 }{ Eur. Phys. J. C
  {\bf 81}, p.~742, (2021)}, arXiv:
  \href{https://arxiv.org/abs/2108.00039}{2108.00039}.

\bibitem{Ingarden96}
R. Ingarden, ``On physical applications of Finsler geometry'', in
  \href{https://doi.org/10.1090/conm/196}{\emph{Finsler Geometry}, Bao, D and
  Chern, S.-S. and Shen, Z. (Editors), Contemporary Mathematics {\bf 196}, AMS,
  1996}.

\bibitem{DuvalHH07}
C.~Duval, Z.~Horvath, P.~Horv{\'a}thy, ``{Geometrical spinoptics and the
  optical Hall effect}'',
  \href{https://dx.doi.org/10.1016/j.geomphys.2006.07.003 }{ J. Geom. Phys.
  {\bf 57}, p.~925, (2007)}, arXiv:
  \href{https://arxiv.org/abs/math-ph/0509031}{math-ph/0509031}.

\bibitem{Duval08}
C.~Duval, ``{Finsler spinoptics}'',
  \href{https://dx.doi.org/10.1007/s00220-008-0573-7 }{ Commun. Math. Phys.
  {\bf 283}, p.~701, (2008)}, arXiv:
  \href{https://arxiv.org/abs/0707.0200}{0707.0200}.

\bibitem{PfeiferW11b}
C.~Pfeifer, M.~N.~R. Wohlfarth, ``{Finsler geometric extension of Einstein
  gravity}'', \href{https://dx.doi.org/10.1103/PhysRevD.85.064009 }{ Phys. Rev.
  D {\bf 85}, p.~064009, (2012)}, arXiv:
  \href{https://arxiv.org/abs/1112.5641}{1112.5641}.

\bibitem{HohmannPV20}
M.~Hohmann, C.~Pfeifer, N.~Voicu, ``{Cosmological Finsler Spacetimes}'',
  \href{https://dx.doi.org/10.3390/universe6050065 }{ Universe {\bf 6}, p.~65,
  (2020)}, arXiv: \href{https://arxiv.org/abs/2003.02299}{2003.02299}.

\bibitem{LiC14}
X.~Li, Z.~Chang, ``{Exact solution of vacuum field equation in Finsler
  spacetime}'', \href{https://dx.doi.org/10.1016/j.physletb.2008.09.010 }{
  Phys. Rev. D {\bf 90}, p.~064049, (2014)}.

\bibitem{Shen23}
B.~Shen, ``{Gravitational lensing effect in the Universe with a Finsler
  background}'', \href{https://dx.doi.org/10.1016/j.geomphys.2023.104999 }{ J.
  Geom. Phys. {\bf 194}, p.~104999, (2023)}.

\bibitem{BaoCS00}
D.~Bao, S.-S. Chern, Z.~Shen, {\em An Introduction to Riemann-Finsler
  Geometry}.
\newblock GTM 200, Springer, New York, 2000, doi:
  \href{https://doi.org/10.1007/978-1-4612-1268-3}{10.1007/978-1-4612-1268-3}.

\bibitem{ShenS16}
Y.-B. Shen, Z.-M. Shen, {\em Introduction to Modern Finsler Geometry}.
\newblock Higher Education Press, World Scientific Publishing, 2016, doi:
  \href{https://doi.org/10.1142/9726}{10.1142/9726}.

\bibitem{BaoCS96}
D. Bao, S.-S. Chern, Z. Shen, ``On the Gauss-Bonnet integrand for 4-dimensional
  Landsberg spaces'', in \href{https://doi.org/10.1090/conm/196}{\emph{Finsler
  Geometry}, Bao, D and Chern, S.-S. and Shen, Z. (Editors), Contemporary
  Mathematics {\bf 196}, AMS, 1996}.

\bibitem{YanoK66}
K.~Yano, S.~Kobayashi, ``{Prolongations of tensor fields and connections to
  tangent bundles I}'', \href{https://dx.doi.org/10.2969/jmsj/01820194 }{ J.
  Math. Soc. Japan {\bf 18}, p.~194, (1966)}.

\bibitem{Davies39}
E.~T. Davies, ``{Lie derivation in generalized metric spaces}'',
  \href{https://dx.doi.org/10.1007/BF02413774 }{ Annali Di matematica {\bf 18},
  p.~261, (1939)}.

\bibitem{EinsteinIH38}
A.~Einstein, L.~Infeld, B.~Hoffmann, ``{The Gravitational equations and the
  problem of motion}'', \href{https://dx.doi.org/10.2307/1968714 }{ Annals
  Math. {\bf 39}, p.~65, (1938)}.

\bibitem{Souriau74}
J.-M. Souriau, ``Mod\`ele de particule \`a spin dans le champ
  \'electromagn\'etique et gravitationnel'', Ann. Inst. Henri Poincar\'e A {\bf
  20}, p.~315 (1974), \url{http://www.numdam.org/item/AIHPA_1974__20_4_315_0}.

\bibitem{Ebin70}
D.~Ebin, ``The manifold of Riemannian metrics'', In Proceedings of symposia in
  pure mathematics, AMS {\bf 15}, p.~11 (1970).

\bibitem{Singer78}
I.~M. Singer, ``{Some Remarks on the Gribov Ambiguity}'',
  \href{https://dx.doi.org/10.1007/BF01609471 }{ Commun. Math. Phys. {\bf 60},
  p.~7, (1978)}.

\bibitem{Francois24}
J.~T. François~André, ``The dressing field method for diffeomorphisms: a
  relational framework'', \href{https://dx.doi.org/10.1088/1751-8121/ad5cad }{
  Journal of Physics A: Mathematical and Theoretical {\bf 57}, p.~305203,
  (2024)}, arXiv: \href{https://arxiv.org/abs/2310.14472}{2310.14472}.

\bibitem{FrancoisR25}
J.~T. Fran\c{c}ois, L.~Ravera, ``{Geometric Relational Framework for
  General-Relativistic Gauge Field Theories}'',
  \href{https://dx.doi.org/10.1002/prop.202400149 }{ Fortsch. Phys. {\bf 73},
  p.~2400149, (2025)}, arXiv:
  \href{https://arxiv.org/abs/2407.04043}{2407.04043}.

\bibitem{Mathisson10}
M.~Mathisson, ``{Republication of: New mechanics of material systems}'',
  \href{https://dx.doi.org/10.1007/s10714-010-0939-y }{ Gen. Rel. Grav. {\bf
  42}, p.~1011, (2010)}.

\bibitem{Dixon15}
W.~G. Dixon, ``{The New Mechanics of Myron Mathisson and Its Subsequent
  Development}'', \href{https://dx.doi.org/10.1007/978-3-319-18335-0_1 }{ In
  Equations of Motion in Relativistic Gravity. Fundamental Theories of Physics,
  Springer, Cham {\bf 179}, p.~1, (2015)}.

\bibitem{Souriau70b}
J.-M. Souriau, ``Sur le mouvement des particules à spin en relativité
  générale'', C. R. Acad. Sci. Paris Sér. A {\bf 271}, p.~751 (1970).

\bibitem{DamourI24}
T.~Damour, P.~Iglesias-Zemmour, ``{Editorial note to: On the motion of spinning
  particles in general relativity by Jean-Marie Souriau}'',
  \href{https://dx.doi.org/10.1007/s10714-024-03294-w }{ Gen. Rel. Grav. {\bf
  56}, p.~127, (2024)}, arXiv:
  \href{https://arxiv.org/abs/2401.10013}{2401.10013}.

\bibitem{Iglesias19}
P.~Iglesias-Zemmour, ``Refraction and reflexion according to the principle of
  general covariance'',
  \href{https://dx.doi.org/https://doi.org/10.1016/j.geomphys.2019.03.013 }{
  Journal of Geometry and Physics {\bf 142}, p.~1, (2019)}, HAL:
  \href{https://hal.science/hal-02401456v1}{hal-02401456v1}.

\bibitem{HohmannPV22}
M.~Hohmann, C.~Pfeifer, N.~Voicu, ``{Mathematical foundations for field
  theories on Finsler spacetimes}'', \href{https://dx.doi.org/10.1063/5.0065944
  }{ J. Math. Phys. {\bf 63}, p.~032503, (2022)}, arXiv:
  \href{https://arxiv.org/abs/2106.14965}{2106.14965}.

\bibitem{Rund59}
H.~Rund, {\em The Differential Geometry of Finsler Spaces}.
\newblock Springer Berlin, Heidelberg, 1959, doi:
  \href{https://doi.org/10.1007/978-3-642-51610-8}{10.1007/978-3-642-51610-8}.

\bibitem{CorinaldesiP51}
E.~Corinaldesi, A.~Papapetrou, ``{Spinning test particles in general
  relativity. 2.}'', \href{https://dx.doi.org/10.1098/rspa.1951.0201 }{ Proc.
  Roy. Soc. Lond. A {\bf 209}, p.~259, (1951)}.

\bibitem{Souriau70}
J.-M. Souriau, {\em Structure des syst\`emes dynamiques}.
\newblock Dunod, Paris, 1970, doi:
  \href{https://doi.org/10.1007/978-1-4612-0281-3}{10.1007/978-1-4612-0281-3}.
\newblock Translation to English: \emph{Structure of Dynamical Systems. A
  Symplectic View of Physics.} (Birkh\"auser, Basel, 1997).

\bibitem{StoneDZ14}
M.~Stone, V.~Dwivedi, T.~Zhou, ``{Berry Phase, Lorentz Covariance, and
  Anomalous Velocity for Dirac and Weyl Particles}'',
  \href{https://dx.doi.org/10.1103/PhysRevD.91.025004 }{ Phys. Rev. D {\bf 91},
  p.~025004, (2015)}, arXiv: \href{https://arxiv.org/abs/1406.0354}{1406.0354}.

\bibitem{HarteO22}
A.~I. Harte, M.~A. Oancea, ``{Spin Hall effects and the localization of
  massless spinning particles}'',
  \href{https://dx.doi.org/10.1103/PhysRevD.105.104061 }{ Phys. Rev. D {\bf
  105}, p.~104061, (2022)}, arXiv:
  \href{https://arxiv.org/abs/2203.01753}{2203.01753}.

\bibitem{Duval72}
C.~Duval, {\em Un modèle de particule à spin dans un champ
  électromagnétique et gravitationnel extérieur}.
\newblock PhD thesis, 3ème cycle, Université de Provence (1972).

\bibitem{Saturnini76}
P.~Saturnini, {\em Un mod\`ele de particules \`a spin de masse nulle dans le
  champ de gravitation}.
\newblock PhD thesis, Universit\'e de Provence (1976).
\newblock HAL: \href{https://hal.science/tel-01344863}{tel-01344863}.

\end{thebibliography}
\end{document}